\documentstyle [12pt, epsfig]{article}
\begin{document}

\hfill {CUMQ/HEP 121}

\hfill {\today}

\vskip 0.5in   \baselineskip 24pt

{\Large  \bigskip
 \centerline{ {\Large $B \rightarrow X_s l^+l^-$ in the left-right
supersymmetric model} }}

\vskip .6in
\def\bar{\overline}

\centerline{Mariana Frank \footnote{Email: mfrank@vax2.concordia.ca}
and Shuquan Nie \footnote{Email: sxnie@alcor.concordia.ca}}
\bigskip
\centerline {\it Department of Physics, Concordia University, 1455 De
Maisonneuve Blvd. W.}
\centerline {\it Montreal, Quebec, Canada, H3G 1M8}

\vskip 0.5in

{\narrower\narrower We analyze the FCNC semileptonic decay $B
\rightarrow X_s l^+l^-$ in a fully left-right supersymmetric
model. We give explicit expressions for all the amplitudes involved in
the process, and compare the numerical results with experimental bounds, in
both the constrained case (where the only flavor violation comes from the
Cabibbo-Kobayashi-Maskawa matrix) and the unconstrained case
(including soft breaking  supersymmetry terms). 
Stringent constraints on the parameter space of the model are obtained. 
We also include constraints from $b \rightarrow s \gamma$.

PACS number(s): 12.60.Jv, 13.20.He, 13.20.-v

\newpage

\section{Introduction}

Flavor changing neutral currents (FCNC) and charge parity (CP) violating
phenomena are some of the best probes for physics beyond the Standard
Model (SM). All existing measurements so far are consistent with the
SM picture involving the Cabibbo-Kobayash-Maskawa (CKM) matrix as the
only source of flavor violation. In the SM, FCNC are absent at tree level,
appear at one loop, but they are effectively suppressed by the GIM
mechanism and
small CKM angles. In supersymmetry, there is no similar mechanism to
suppress the loop contributions to either flavor or CP violating
phenomena. Experimental studies of flavor physics, especially in B
decays, appear essential for the understanding of the mechanism for
supersymmetry breaking. With increased statistical power of
experiments at B factories, rare B decays will be measured very
precisely.

In this paper we investigate the relevance of new physics in the
semileptonic inclusive decay   $B \rightarrow X_s l^+l^-$ in a fully
left-right supersymmetric model. Investigation of the process $b
\rightarrow s \gamma $ in this model has shown distinctive signs
from the minimal
supersymmetric standard model (MSSM) scenario \cite{fn}. An analysis of $B
\rightarrow X_s l^+l^-$ would provide some complementary information. The
semileptonic decay $B \rightarrow X_s l^+l^-$ is a benchmark of
charmless $b$-decays
with strange particles in the final state. The process is
experimentally clean, but the
expected SM branching ratio in the $10^{-6}-10^{-5}$ region makes it not easily
detectable at B-factories. Therefore it provides for excellent opportunities to
test physics beyond the SM. It also offers more detailed information about the
flavor structure of the model, it provides a good test of the structure of the
$Zbs$ vertex, making it particularly well suited to distinguish the left-right
symmetry over the MSSM.

Experimentally, BELLE has recently announced the first evidence for
the exclusive process $B \rightarrow K^{*} l^+l^-$. BABAR and
BELLE have upper
bounds for $B \rightarrow (K, K^{*})+(e^+e^-, \mu^+ \mu^-)$ which are
very close to
the SM estimates. Experimentally, the bounds on the branching ratios
are \cite{BELLE}
\begin{eqnarray}
BR(B \rightarrow X_s \mu^+ \mu^-) &<& 19.1 \times 10^{-6} \ @~90\% \
\mathrm{C.L.},
\nonumber\\
BR(B \rightarrow X_s e^+ e^-) &<& 10.1 \times 10^{-6} \ @~90\% \ \mathrm{C.L.},
\end{eqnarray}
typically a factor of 3 away from the SM estimates, where the
next-to-next-to leading logarithmic (NNLO) calculations to ${\cal
O}(1/\alpha_s)$ \cite{NNLO} have appeared recently
\begin{eqnarray}
BR(B \rightarrow X_s \mu^+ \mu^-) <(4.15 \pm 0.70) \times 10^{-6},  \nonumber\\
BR(B \rightarrow X_s e^+ e^-) <(6.89 \pm 1.01) \times 10^{-6}.
\end{eqnarray}
Here we restrict ourselves to the analysis of inclusive processes only: in
the case of exclusive
decay rates, hadronic matrix element uncertainties obscure model predictions.

Semileptonic charmless B decays have been studied previously by many authors
in the framework of supersymmetric models with a universal soft
supersymmetry breaking terms
\cite{bsll}. Recently an analysis of SUSY models with
non-universal soft breaking terms at
the grand unification scale has appeared in Ref. \cite{GK}. Although
attempts have been
made to reconcile $b \rightarrow s \gamma$ with right-handed $b$-quark
decays \cite{gronau}, a complete analysis of $ B \rightarrow X_s l^+
l^-$ for a fully
left-right supersymmetric model is still lacking.

The Left-Right Supersymmetric (LRSUSY) models \cite{history, frank1}, based
on the symmetry group $SU(2)_L
\times SU(2)_R \times U(1)_{B-L}$, incorporate the advantages of supersymmetry
with a natural framework for allowing neutrino masses through the
seesaw mechanism
\cite{mohapatra}. Various other scenarios incorporate some forms of the left-right
symmetry within supersymmetry.  LRSUSY models have the attractive
feature that
they can be embedded in a supersymmetric grand unified theory such as
$SO(10)$ \cite{SO10}, while not bound by lepton quark unification. They would also appear in model building realistic brane
worlds from Type I strings. This involves the left-right
supersymmetry, with supersymmetry broken either at the string scale
$M_{SUSY} \approx 10^{10-12}$ GeV, or at $M_{SUSY} \approx 1$ TeV, the
difference having implications for the gauge unification \cite{string}.

In this paper we study all contributions of the LRSUSY model to the
branching ratio and the asymmetry of $B \rightarrow X_s l^+l^-$ at one-loop level. The decay can be
mediated by the left-handed and right-handed W and Z bosons, and by charged Higgs
bosons as in the nonsupersymmetric case, but also by
charginos, neutralinos and gluinos. The structure of the LRSUSY model
provides a significant contributions
to the decay $B \rightarrow X_s l^+l^-$ from the right-handed squarks
and an enlarged
gaugino-Higgsino sector with right-handed couplings, which is not as
constrained as the
right-handed gauge sector in the LRSUSY model. We anticipate
that these could contribute a large enhancement of the decay rate
and would constrain some of the parameters of the model.

The paper is organized as follows. We describe the structure of the model
in Sec. II, with
particular emphasis on the gaugino-Higgsino and squark structure. In
Sec. III, we give the
supersymmetric contributions in the LRSUSY model to the decay
$B \rightarrow X_s l^+l^-$. We confront the calculation with
experimental results in Sec. IV, where we present the numerical analysis to
constrain the
parameters of the model for two scenarios: one with the CKM flavor mixing
only, the other
including supersymmetric soft breaking flavor violation terms. We reach our
conclusions in Sec. V.

\section{The Model}

The LRSUSY electroweak symmetry group, $SU(2)_{L}\times
SU(2)_{R}\times U(1)_{B-L}$,
has matter
doublets for both left- and right-handed fermions and their
corresponding left-
and right-handed scalar partners (sleptons and squarks)~\cite{frank1}.
In the gauge sector,
corresponding to $SU(2)_{L}$ and $SU(2)_{R}$, there are triplet
gauge bosons $(W^{+}, W^{-},W^{0})_{L}$, $(W^{+}, W^{-},W^{0})_{R}$,
respectively, and a singlet
gauge boson $V$ corresponding to $U(1)_{B-L}$, together with their
superpartners. The Higgs sector of this model
consists of two Higgs bi-doublets, $\Phi_{u}(\frac{1}{2},\frac{1}{2},0)$
and $\Phi_{d}(\frac{1}{2},\frac{1}{2},0)$, which are required to give masses
to the up and down quarks.  The spontaneous symmetry breaking of the group
$SU(2)_{R}\times U(1)_{B-L}$ to the hypercharge symmetry group
$U(1)_{Y}$ is
accomplished by giving vacuum expectation values to a pair of Higgs
triplet fields
$\Delta_{L}(1,0,2)$  and $\Delta_{R}(0,1,2)$, which transform as the
adjoint
representation of $SU(2)_R$. The choice of two triplets (versus four
doublets) is
preferred because with this choice a large Majorana mass can be
generated (through
the see-saw mechanism) for the right-handed neutrino and a small one for
the left-handed neutrino~\cite{mohapatra}.
In addition to the triplets $\Delta_{L,R}$, the model must contain two
additional triplets, $\delta_{L}(1,0,-2)$ and $\delta_{R}(0,1,-2)$, with
quantum number $B-L= -2$, to insure cancellation of the anomalies which
would otherwise occur in the fermionic sector.
The superpotential for the LRSUSY model is
\begin{eqnarray}
\label{superpotential}
W_{LRSUSY} & = & {\bf h}_{q}^{(i)} Q^T\tau_{2}\Phi_{i} \tau_{2}Q^{c} + {\bf
h}_{l}^{(i)}
L^T\tau_{2}\Phi_{i} \tau_{2}L^{c} + i({\bf h}_{LR}L^T\tau_{2} \Delta_L L
 \nonumber \\
&+ & {\bf
h}_{LR}L^{cT}\tau_{2}
\Delta_R L^{c}) + M_{LR}\left [Tr (\Delta_L  \delta_L +\Delta_R
\delta_R)\right] \nonumber \\
&+& \mu_{ij}Tr(\tau_{2}\Phi^{T}_{i} \tau_{2} \Phi_{j}) +W_{NR},
\end{eqnarray}
where $W_{NR}$ denotes (possible) non-renormalizable terms arising
from higher scale physics or Planck scale effects~\cite{recmohapatra}.
The presence of these terms
insures that, when the SUSY breaking scale is above $M_{W_{R}}$, the
ground state is R-parity conserving~\cite{km}.

The neutral Higgs fields acquire non-zero vacuum
expectation values $(VEV's)$ through spontaneous symmetry breaking
\begin{eqnarray}
\langle \Delta \rangle_{L,R} = \left(\begin{array}{cc}
0&0\\v_{L,R}&0
\end{array}\right),
~\rm{and}~
\langle \Phi \rangle_{u,d} = \left (\begin{array}{cc}
\kappa_{u,d}&0\\0&\kappa^{\prime}_{u,d} e^{i\omega}
\end{array}\right).
\nonumber
\end{eqnarray}
$\langle \Phi \rangle_{u,d}$ cause the mixing of $W_{L}$ and $W_{R}$
bosons with $CP$-violating
phase $\omega$, which is set to zero in the analysis. The non-zero Higgs $VEV's$
break both parity and $SU(2)_{R}$.
In the first stage of breaking, the right-handed gauge bosons, $W_{R}$ and
$Z_{R}$ acquire masses proportional to $v_{R}$ and become much heavier
than the SM (left-handed) gauge bosons $W_{L}$ and $Z_{L}$, which pick
up masses proportional to $\kappa_{u}$ and $\kappa_{d}$ at the second stage of
breaking.

In the supersymmetric sector of the model there are six singly-charged
charginos, corresponding to $\tilde\lambda_{L}$,
$\tilde\lambda_{R}$, $\tilde\phi_{u}$,
$\tilde\phi_{d}$, $\tilde\Delta_{L}^{\pm}$, and
$\tilde\Delta_{R}^{\pm}$.
The model also has eleven neutralinos, corresponding to
$\tilde\lambda_{Z}$,
$\tilde\lambda_{Z^{\prime}}$,
$\tilde\lambda_{V}$,  $\tilde\phi_{u1}^0$, $\tilde\phi_{u2}^0$,
$\tilde\phi_{d1}^0$,  $\tilde\phi_{d2}^0$, $\tilde\Delta_{L}^0$,
$\tilde\Delta_{R}^0$,  $\tilde\delta_{L}^0$, and
$\tilde\delta_{R}^0$. Although $\Delta_{L}$
is not necessary for symmetry breaking~\cite{huitu}, and is
introduced only for preserving the left-right symmetry, both
$\Delta_{L}^{--}({\tilde \Delta_{L}^{--}})$ and its
right-handed counterparts $\Delta_{R}^{--}({\tilde \Delta_{R}^{--}})$
play very important
roles in lepton phenomenology of the LRSUSY model. The doubly charged Higgs
and Higgsinos do not
affect quark phenomenology, but the neutral and singly charged
components do, through
mixings in the chargino and neutralino mass matrices. We include only
the ${\tilde \Delta}_R$ contribution in the numerical analysis.

The supersymmetric sources of flavor violation in the LRSUSY model
come from either the Yukawa potential or the trilinear scalar couplings.
The interactions of fermions with scalar (Higgs) fields have the following
form
\begin{eqnarray}
\label{eq:yukawa}
{\cal L}_Y &=& {\bf h}_u\bar{Q}_L \Phi_u Q_R + {\bf h}_d \bar{Q}_L \Phi_d
Q_R\,
+{\bf h}_\nu\bar{L}_L  \Phi_u L_R + {\bf h}_e \bar{L}_L \Phi_d
L_R+\,H.c.,\nonumber \\
{\cal L}_M &=& i{\bf h}_{LR}(L_L^TC^{-1}\tau_2\Delta_LL_L+
L_R^TC^{-1}\tau_2\Delta_RL_R) +\,H.c.,
\end{eqnarray}
where ${\bf h}_u$, ${\bf h}_d$, ${\bf h}_{\nu}$ and ${\bf h}_e$ are the Yukawa
couplings for the up and down quarks and neutrino and electron, respectively,
and ${\bf h}_{LR}$ is the coupling for the triplet Higgs bosons.  The left-right symmetry
requires all ${\bf h}$-matrices to be Hermitean in the generation space and
${\bf h}_{LR}$ matrix to be symmetric.
In the universal case, there is no intergenerational mixings for
squarks and the only source of flavor mixing comes from the CKM
matrix. We will analyze this
case first. Next we will look at the case in which intergenerational mixings in the squark
sector are permitted and consider
the effect of intergenerational mixings on the rate of the process $B
\rightarrow X_s l^+l^-$.

\section{The analytic formulas}

The effective Hamiltonian for the decay $B \rightarrow X_s l^+l^-$ at the scale
$\mu$ in the LRSUSY model is given by
\begin{equation}
{\cal H}_{eff}=-\frac{4 G_F}{\sqrt{2}}K_{tb}K^*_{ts} \sum_i [ C_i(\mu)
Q_i(\mu) + {\tilde C}_i(\mu) {\tilde Q}_i(\mu)].
\end{equation}
The operators relevant to the process $b \rightarrow s l^+l^-$ in the LRSUSY model
are
\begin{eqnarray}
Q_{7}&=&m_b (q_{\nu}/q^2) (\bar{s} i\sigma_{\mu \nu} P_R b)( \bar{l}
\gamma_{\mu}l),
\nonumber \\
{\tilde Q}_{7}&=&m_b (q_{\nu}/q^2) (\bar{s} i\sigma_{\mu \nu} P_L b)(
\bar{l} \gamma_{\mu}l),
\nonumber \\
Q_{9}&=&( \bar{s}\gamma_{\mu}P_L b) (\bar{l} \gamma_{\mu} l), \nonumber \\
{\tilde Q}_{9}&=&( \bar{s}\gamma_{\mu}P_R b) (\bar{l} \gamma_{\mu}
l), \nonumber \\
Q_{10}&=&( \bar{s}\gamma_{\mu}P_L b) (\bar{l} \gamma_{\mu}\gamma_5
l), \nonumber \\
{\tilde Q}_{10}&=&( \bar{s}\gamma_{\mu}P_R b) (\bar{l} \gamma_{\mu}\gamma_5 l).
\end{eqnarray}
The Wilson coefficients $C_{i}$ and ${\tilde C}_i$ are initially evaluated
at the electroweak or soft supersymmetry breaking scale, then evolved down
to the scale $\mu$. In the SM and constrained SUSY models, ${\tilde
Q}_i$ contributions
are generally suppressed by ${\cal O}(m_s/m_b)$ compared with the
contributions from $Q_i$.
However this is not the case in generic SUSY models such as
non-universal models. In Ref. \cite{GK}
the operator ${\tilde Q}_7$ was included in the analysis. Due to the
left-right symmetry,
we are motivated to consider all contributions from both chirality operators.

The decay $b \rightarrow s l^+ l^-$ can be mediated by either the photon or the $Z_L$, $Z_R$
bosons, or it can proceed through the box diagrams.
As in the MSSM, the $Z_L$ boson contributions dominate where there is explicit
$SU(2)_L$ symmetry breaking, and the $Z_R$ boson contributions are important where
there is explicit
$SU(2)_R$ symmetry breaking, i.e., both cases in which left and right
squarks occur
in the same loop diagram. In these cases, the $Z$ diagrams are enhanced
by $m_{\tilde q}^2/M_Z^2$ with respect to the photon graphs.
This could be an order of magnitude for the regular $Z_L$ boson, but only of order 1 for the $Z_R$ boson. We describe these contributions in detail below.

\subsection{The photon monopole and penguin graphs}

\begin{figure}
\centerline{ \epsfysize 2.0in
\rotatebox{360}{\epsfbox{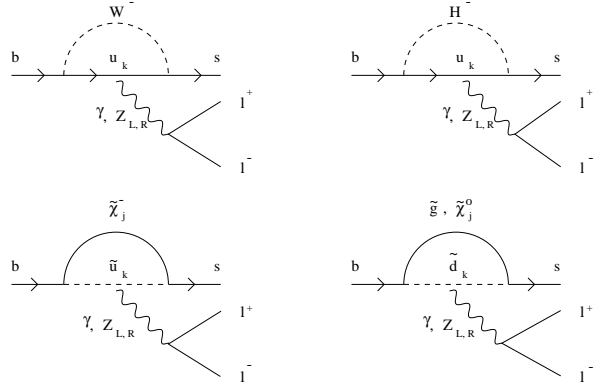}}  }
\caption{Penguin diagrams which induce the decay $b \rightarrow s
l^+ l^-$ in the LRSUSY model. The outgoing photon and Z boson lines can be
attached in all possible ways. }
\protect \label{pole}
\end{figure}

The penguin monopole and dipole graphs are presented in Fig. \ref{pole}.
We first give the contributions for the constrained model, i.e.,
assuming the CKM matrix
is the only source of flavor violation. The left-handed monopole
contributions are given by
\begin{eqnarray}
A_{SM}^{LL}&=&-\frac{\alpha_W \alpha}{2} \frac{1}{M_{W_L}^2}K_{ts}^{\ast}K_{tb}
\{x_{tW}[Q_uF_7(x_{tW})+F_8(x_{tW})]\nonumber \\
& & + \frac23 Q_u[\frac{\ln(x_{tW})}{x_{tW}-1} -1
+\ln(\frac{m_c^2}{M_{W_L}^2}) +f(\frac{q^2}{m_b^2})]\},  \\
A_{H^-}^{LL}&=&-\frac{\alpha_W \alpha}{2} \frac{1}{M_{W_L}^2}K_{ts}^{\ast}K_{tb}
x_{tH} \cot ^2 \beta[Q_u F_5(x_{tH})-F_6(x_{tH})],  \\
A_{{\tilde \chi}^-}^{LL}&=&- \alpha_W \alpha \sum_{j=1}^{5} \sum_{k=1}^{6}
\frac{1}{m_{{\tilde u}_k}^2} (G_{UL}^{jkb}- H_{UR}^{jkb})(G_{UL}^{\ast jks}-
H_{UR}^{\ast jks})\nonumber \\
&  & \times [Q_u F_6(x_{\tilde{\chi}_j^- \tilde{u}_k})-F_5(x_{\tilde{\chi}_j^-
\tilde{u}_k})],  \\
A_{\tilde g}^{LL}&=&-2 \alpha_s \alpha C(R) Q_d \sum_{k=1}^6
\frac{1}{m_{{\tilde
d}_k}^2} \Gamma_{DL}^{kb} \Gamma_{DL}^{\ast ks} F_6(x_{\tilde g \tilde{d}_k}),
\\
A_{\tilde \chi^0}^{LL}&=&- \alpha_W \alpha Q_d \sum_{j=1}^9 \sum_{k=1}^6
\frac{1}{m_{{\tilde d}_k}^2} ( \sqrt{2} G_{0DL}^{jkb}-H_{0DR}^{jkb}) (
\sqrt{2}G_{0DL}^{\ast jks}-H_{0DR}^{\ast jks}) F_6(x_{\tilde \chi^0_j \tilde{d}_k}).
\end{eqnarray}
The monopole contributions with $W_L$ replaced by $W_R$, as in the left-right symmetric model (LRM), are suppressed by $\frac{M_{W_L}^2}{M_{W_R}^2}$ and thus negligible.
The only RR monopole contributions then come from the supersymmetric sector
\begin{eqnarray}
A_{{\tilde \chi}^-}^{RR}&=&- \alpha_W \alpha \sum_{j=1}^{5} \sum_{k=1}^{6}
\frac{1}{m_{{\tilde u}_k}^2} (G_{UR}^{jkb}- H_{UL}^{jkb})(G_{UR}^{\ast jks}-
H_{UL}^{\ast jks})\nonumber \\
& & \times [Q_uF_6(x_{\tilde \chi_j^- \tilde{u}_k})-F_5(x_{\tilde
\chi_j^- \tilde{u}_k})], \\
A_{\tilde g}^{RR}&=&- 2\alpha_s \alpha C(R) Q_d \sum_{k=1}^6
\frac{1}{m_{{\tilde
d}_k}^2} \Gamma_{DR}^{kb} \Gamma_{DR}^{\ast ks} F_6(x_{\tilde g \tilde{d}_k}),
\\
A_{\tilde \chi^0}^{RR}&=&- \alpha_W \alpha Q_d \sum_{j=1}^9 \sum_{k=1}^6
\frac{1}{m_{{\tilde d}_k}^2} ( \sqrt{2}G_{0DR}^{jkb}-H_{0DL}^{jkb}) (
\sqrt{2}G_{0DR}^{\ast jks}-H_{0DL}^{\ast jks}) F_6(x_{\tilde \chi^0_j \tilde{d}_k}).
\end{eqnarray}
The dipole LR and RL contributions can be obtained by multiplying the
contributions to the decay $b \rightarrow s \gamma$ by the factor $\sqrt{4 \pi
\alpha}$
\begin{eqnarray}
A_{SM}^{LR} &=& \frac{\alpha \alpha_W}{2} \frac{1}{M_{W_L}^2} K^*_{ts} K_{tb} 3 x_{tW} [ Q_u F_1 (x_{tW})
+F_2(x_{tW})],  \\
A_{H^-}^{LR} &=& \frac{\alpha \alpha_W}{2} \frac{1}{M_{W_L}^2} K^*_{ts} K_{tb}  x_{tH} \{
\cot^2 \beta [ Q_u F_1 (x_{tH})+F_2(x_{tH})] \nonumber \\
& &+[ Q_u F_3 (x_{tH})+F_4 (x_{tH})]  \},  \\
A_{\tilde{g}}^{LR} &=& - 2 \alpha \alpha_s Q_d C(R)
\sum_{k=1}^6
\frac{1}{m_{\tilde{d}_k}^2} \{ \Gamma_{DL}^{kb} \Gamma_{DL}^{*ks}
F_2(x_{\tilde{g} \tilde{d_k}})
-\frac{m_{\tilde{g}}}{m_b} \Gamma_{DR}^{kb} \Gamma_{DL}^{*ks}
F_4(x_{\tilde{g} \tilde{d_k}}) \}, \\
A_{\tilde{\chi}^-}^{LR} &=& - \alpha \alpha_W
\sum_{j=1}^5 \sum_{k=1}^6
\frac{1}{m_{\tilde{u}_k}^2} \{
(G_{UL}^{jkb}-H_{UR}^{jkb})(G_{UL}^{*jks}-H_{UR}^{*jks})
[ F_1(x_{\tilde{\chi}_j^- \tilde{u}_k})+ Q_u F_2(x_{\tilde{\chi}_j^-
\tilde{u}_k})] \nonumber \\
      & & +\frac{m_{\tilde{\chi}_j^-}}{m_b} (G_{UR}^{jkb}-H_{UL}^{jkb})
(G_{UL}^{*jks}
-H_{UR}^{*jks}) [F_3 (x_{\tilde{\chi}_j^- \tilde{u}_k})+ Q_u
F_4(x_{\tilde{\chi}_j^- \tilde{u}_k})] \}, \\
A_{\tilde{\chi}^0}^{LR} &=& - \alpha \alpha_W Q_d
\sum_{j=1}^9 \sum_{k=1}^6
\frac{1}{m_{\tilde{d}_k}^2} \{
(\sqrt{2}G_{0DL}^{jkb}-H_{0DR}^{jkb})(\sqrt{2}G_{0DL}^{*jks}-H_{0DR}^{*jks})
F_2(x_{\tilde{\chi}_j^0 \tilde{d}_k}) \nonumber \\
      & & +\frac{m_{\tilde{\chi}_j^0}}{m_b}
(\sqrt{2}G_{0DR}^{jkb}-H_{0DL}^{jkb}) (\sqrt{2} G_{0DL}^{*jks}
-H_{0DR}^{*jks})  F_4(x_{\tilde{\chi}_j^0 \tilde{d}_k}) \},
\end{eqnarray}
and
\begin{eqnarray}
A_{\tilde{g}}^{RL} &=& - 2 \alpha \alpha_s Q_d C(R)
\sum_{k=1}^6
\frac{1}{m_{\tilde{d}_k}^2} \{ \Gamma_{DR}^{kb} \Gamma_{DR}^{*ks}
F_2(x_{\tilde{g} \tilde{d_k}})
-\frac{m_{\tilde{g}}}{m_b} \Gamma_{DL}^{kb} \Gamma_{DR}^{*ks}
F_4(x_{\tilde{g} \tilde{d_k}}) \}, \\
A_{\tilde{\chi}^-}^{RL} &=& -  \alpha \alpha_W
\sum_{j=1}^5 \sum_{k=1}^6
\frac{1}{m_{\tilde{u}_k}^2} \{
(G_{UR}^{jkb}-H_{UL}^{jkb})(G_{UR}^{*jks}-H_{UL}^{*jks})
[ F_1(x_{\tilde{\chi}_j^- \tilde{u}_k})+ Q_u F_2(x_{\tilde{\chi}_j^-
\tilde{u}_k})] \nonumber \\
      & & +\frac{m_{\tilde{\chi}_j^-}}{m_b} (G_{UL}^{jkb}-H_{UR}^{jkb})
(G_{UR}^{*jks}
-H_{UL}^{*jks}) [F_3 (x_{\tilde{\chi}_j^- \tilde{u}_k})+ Q_u
F_4(x_{\tilde{\chi}_j^- \tilde{u}_k})] \}, \\
A_{\tilde{\chi}^0}^{RL} &=& -  \alpha \alpha_W Q_d
\sum_{j=1}^9 \sum_{k=1}^6
\frac{1}{m_{\tilde{d}_k}^2} \{
(\sqrt{2}G_{0DR}^{jkb}-H_{0DL}^{jkb})(\sqrt{2}G_{0DR}^{*jks}-H_{0DL}^{*jks})
F_2(x_{\tilde{\chi}_j^0 \tilde{d}_k}) \nonumber \\
      & & +\frac{m_{\tilde{\chi}_j^0}}{m_b}
(\sqrt{2}G_{0DL}^{jkb}-H_{0DR}^{jkb}) (\sqrt{2} G_{0DR}^{*jks}
-H_{0DL}^{*jks})  F_4(x_{\tilde{\chi}_j^0 \tilde{d}_k}) \},
\end{eqnarray}
where vertex mixing matrices $G$, $H$, $G_0$ and $H_0$ are defined in
the Appendix. The convention
$x_{ab}=m_a^2/m_b^2$ is used. $C(R) =4/3$ is the quadratic Casimir
operator of the fundamental representation of $SU(3)_C$.

\subsection{The $Z_L, \ Z_R$ penguin graphs}

The process $b \rightarrow s l^+l^-$ is also induced by the effective
couplings of
the $Z_L$ and $Z_R$. The diagrams are analogous to the photon
graphs where the photon is replaced by the $Z$ propagator.
The amplitudes for the $Z_L$ mediated graphs are
\begin{eqnarray}
A_{SM}^{Z_L} &=&-\frac{\alpha_W^2}{2} \frac {1}{M_{Z_L}^2 \cos^2 \theta_W}
K_{ts}^{\ast}K_{tb} x_{tW} F_9(x_{tW}), \\
A_{H^-}^{Z_L} & = &-\frac{\alpha_W^2}{4} \frac {1}{M_{Z_L}^2 \cos^2 \theta_W}
K_{ts}^{\ast}K_{tb} \cot ^2 \beta x_{tW} x_{tH} [ F_3(x_{tH}) +F_4(x_{tH}) ],\\
A_{\tilde g}^{Z_L} &=&- \alpha_W \alpha_s C(R)\frac {1}{M_{Z_L}^2 \cos^2
\theta_W} \sum_{h,k=1}^6   \Gamma_{DL}^{kb} \Gamma_{DL}^{\ast hs}
\sum_{m=1}^3
\Gamma_{DR}^{hm} \Gamma_{DR}^{\ast km} G_0(x_{\tilde{d}_k \tilde g},
x_{\tilde{d}_h \tilde g }), \\
A_{{\tilde \chi}^-}^{Z_L} &=&-\frac{ \alpha_W \alpha}{2}\frac
{1}{M_{Z_L}^2 \cos^2 \theta_W} \sum_{h,k=1}^6  \sum_{i, j=1}^5
(G_{UL}^{jkb}-H_{UR}^{jkb})(G_{UL}^{\ast ihs}-H_{UR}^{\ast ihs}) \nonumber \\
&  &\times  \left \{\delta_{ij} \sum_{m=1}^3 \Gamma_{UL}^{hm}
\Gamma_{UL}^{\ast km}  G_0(x_{\tilde{u}_h \tilde \chi_i^- }, x_{\tilde{u}_k \tilde \chi_j^-}) +
\delta_{hk} [2\sqrt{x_{\tilde \chi_j^- \tilde{u}_k} x_{\tilde \chi_i^- \tilde{u}_k}}
F_0(x_{\tilde \chi_j^- \tilde{u}_k}, x_{\tilde \chi_i^- \tilde{u}_k}) \right. \nonumber \\
& & \left. \times U_{i1}U^{\ast}_{j1}
    -G_0(x_{\tilde \chi_j^- \tilde{u}_k}, x_{\tilde \chi_i^- \tilde{u}_k})
V^{\ast}_{i1}V_{j1}] \right \}, \\
A_{{\tilde \chi}^0}^{Z_L} &=&-\frac{ \alpha_W \alpha}{2}\frac
{1}{M_{Z_L}^2 \cos^2 \theta_W} \sum_{h,k=1}^6  \sum_{i, j=1}^9
(\sqrt{2} G_{0DL}^{jkb}-H_{0DR}^{jkb})(\sqrt{2} G_{0DL}^{\ast
ihs}-H_{0DR}^{\ast ihs})
\nonumber \\
& &\times  \left \{\delta_{ij} \sum_{m=1}^3 \Gamma_{DL}^{hm}
\Gamma_{DL}^{\ast km}  G_0(x_{\tilde{d}_h \tilde \chi_i^0 }, x_{\tilde{d}_k \tilde \chi_j^0}) +
\delta_{hk} [2\sqrt{x_{\tilde \chi_j^0 \tilde{d}_k} x_{\tilde \chi_i^0 \tilde{d}_k}}
F_0(x_{\tilde \chi_j^0 \tilde{d}_k}, x_{\tilde \chi_i^0 \tilde{d}_k}) \right. \nonumber \\
& & \left. \times (N_{i4}N^{\ast}_{j4}-N_{i5}N^{\ast}_{j5})
    -G_0(x_{\tilde \chi_j^0 \tilde d_k}, x_{\tilde \chi_i^0 \tilde{d}_k})
(N^{\ast}_{i4}N_{j4}-N^{\ast}_{i5}N_{j5})] \right \},
\end{eqnarray}
where we have included in the expressions both cases in which one vertex is
gaugino and the other Higgsino, and the case in which we have two gaugino
vertices. Similarly we obtain, for the $Z_R$ mediated graphs
\begin{eqnarray}
A_{\tilde g}^{Z_R} &=&- \alpha_W \alpha_s C(R) \frac {\cos 2
\theta_W}{M_{Z_R}^2 \cos^2
\theta_W} \sum_{h,k=1}^6  \Gamma_{DR}^{kb} \Gamma_{DR}^{\ast hs} \sum_{m=1}^3
\Gamma_{DL}^{hm} \Gamma_{DL}^{\ast km} G_0(x_{\tilde{d}_k \tilde g },
x_{\tilde{d}_h \tilde g }), \\
A_{{\tilde \chi}^-}^{Z_R} &=&-\frac{ \alpha_W \alpha}{2}\frac {\cos 2
\theta_W}{M_{Z_R}^2 \cos^2
\theta_W} \sum_{h,k=1}^6  \sum_{i, j=1}^5
(G_{UR}^{jkb}-H_{UL}^{jkb})(G_{UR}^{\ast ihs}-H_{UL}^{\ast ihs}) \nonumber \\
&  & \times  \left \{\delta_{ij} \sum_{m=1}^3 \Gamma_{UR}^{hm}
\Gamma_{UR}^{\ast km}  G_0(x_{\tilde{u}_h \tilde \chi_i^- }, x_{\tilde{u}_k \tilde \chi_j^-}) +
\delta_{hk} [2\sqrt{x_{\tilde \chi_j^- \tilde u_k} x_{\tilde \chi_i^- \tilde u_k}}
F_0(x_{\tilde
\chi_j^- \tilde{u}_k}, x_{\tilde \chi_i^- \tilde{u}_k}) \right. \nonumber \\
& & \left. \times U_{i2}U^{\ast}_{j2}
  -G_0(x_{\tilde \chi_j^- \tilde{u}_k}, x_{\tilde \chi_i^- \tilde{u}_k})
V^{\ast}_{i2}V_{j2}] \right \}, \\
A_{{\tilde \chi}^0}^{Z_R} &=&-\frac{ \alpha_W \alpha}{2}\frac {\cos 2
\theta_W}{M_{Z_R}^2 \cos^2 \theta_W} \sum_{h,k=1}^6  \sum_{i, j=1}^9
(\sqrt{2}G_{0DR}^{jkb}-H_{0DL}^{jkb})(\sqrt{2} G_{0DR}^{\ast
ihs}-H_{0DL}^{\ast ihs})
\nonumber \\
& & \times \left \{\delta_{ij} \sum_{m=1}^3 \Gamma_{DR}^{hm}
\Gamma_{DR}^{\ast km}  G_0(x_{\tilde{d}_h \tilde \chi_i^0 }, x_{\tilde{d}_k \tilde \chi_j^0}) +
\delta_{hk} [2\sqrt{x_{\tilde \chi_j^0 \tilde{d}_k} x_{\tilde \chi_i^0 \tilde{d}_k}}
F_0(x_{\tilde \chi_j^0 \tilde{d}_k}, x_{\tilde \chi_i^0 \tilde{d}_k}) \right. \nonumber \\
& & \left. \times (N_{i4}N^{\ast}_{j4}-N_{i5}N^{\ast}_{j5})
    -G_0(x_{\tilde \chi_j^0 \tilde{d}_k}, x_{\tilde \chi_i^0 \tilde{d}_k})
(N^{\ast}_{i4}N_{j4}-N^{\ast}_{i5}N_{j5})] \right \}.
\end{eqnarray}
As in the MSSM \cite{bsll}, when the $Z_{L,R}$ bosons are exchanged,
gluino- and neutralino-induced contributions to the total amplitudes
are suppressed
by ${\cal O}(q^2/M_{Z_{L,R}}^2)$ with respect to the photon penguin
contributions. In
addition, the $Z_R$ contribution is suppressed with respect to the
left-handed one by ${\cal O}(M_{Z_L}^2/M_{Z_{R}}^2)$.

\subsection{The box diagrams}

\begin{figure}
\centerline{ \epsfysize 2.0in
\rotatebox{360}{\epsfbox{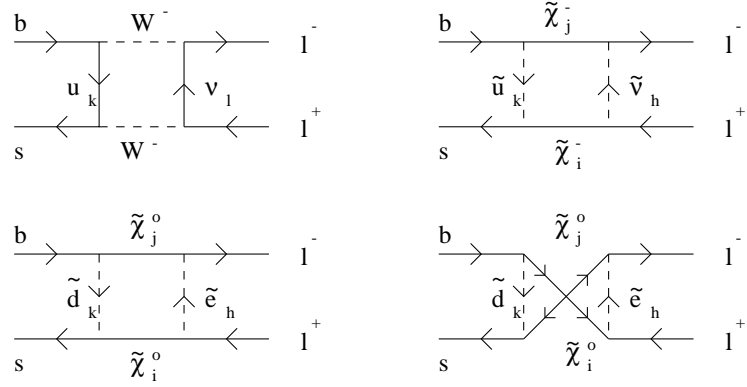}}  }
\caption{Box diagrams contributing to the decay $b \rightarrow s
l^+ l^-$ in the LRSUSY model. Clashing arrows on the fermion lines indicate
a Majorana mass insertion. }
\protect \label{box}
\end{figure}

The box graphs are presented in Fig. \ref{box}. The explicit
contributions are given by
\begin{eqnarray}
A_{SM}^{\Box} &=&-\frac{\alpha_W^2}{4} \frac {1}{M_{W_L}^2}
K_{ts}^{\ast}K_{tb} [G (x_{tW}, 0)- G(0,0)], \\
A_{{\tilde \chi}^-}^{L \Box} &=&\frac {\alpha_W^2}{4}
\sum_{h,k=1}^6  \sum_{i, j=1}^5  \frac {1}{m_{\tilde \chi_i^-}^2}
(G_{UL}^{jkb}-H_{UR}^{jkb})(G_{UL}^{\ast iks}-H_{UR}^{\ast iks}) \nonumber \\
& & \times G_{NL}^{\ast jhl} G_{NL}^{ihl}
    G^{\prime}(x_{\tilde{u}_k \tilde \chi_j^- }, x_{\tilde{\nu}_h \tilde
\chi_j^-}, x_{\tilde \chi_i^- \tilde \chi_j^- }),  \\
A_{{\tilde \chi}^0}^{L \Box} &=& \frac {\alpha_W^2}{2}
  \sum_{h,k=1}^6  \sum_{i, j=1}^9 \frac {1}{m_{\tilde \chi_j^0}^2}
(\sqrt{2}G_{0DL}^{jkb}-H_{0DR}^{jkb})(\sqrt{2}G_{0DL}^{\ast iks}-H_{0DR}^{\ast
iks}) \left [G_{0EL}^{\ast jhl}G_{0EL}^{ ihl} \right. \nonumber \\
& &\left. \times  G^{\prime}(x_{\tilde{d}_k \tilde \chi_j^0}, x_{\tilde l_h
\tilde \chi_j^0}, x_{\tilde \chi_i^0 \tilde \chi_j^0 })
-2 G_{0EL}^{\ast ihl}G_{0EL}^{ jhl}  \sqrt{x_{\tilde \chi_i^0 \tilde \chi_j^0}}
F^{\prime}(x_{\tilde{d}_k \tilde \chi_j^0 }, x_{\tilde l_h
\tilde \chi_j^0}, x_{\tilde \chi_i^0 \tilde \chi_j^0 }) \right ], \\
A_{{\tilde \chi}^0}^{L \Box \prime} &=&-\frac {\alpha_W^2}{2}
  \sum_{h,k=1}^6  \sum_{i, j=1}^9 \frac {1}{m_{\tilde \chi_j^0}^2}
(\sqrt{2}G_{0DL}^{jkb}-H_{0DR}^{jkb})(\sqrt{2}G_{0DL}^{\ast iks}-H_{0DR}^{\ast
iks})\left [G_{0ER}^{\ast jhl}G_{0ER}^{ ihl} \right. \nonumber \\
& &\left. \times  G^{\prime}(x_{\tilde{d}_k \tilde \chi_j^0 }, x_{\tilde l_h
\tilde \chi_j^0}, x_{\tilde \chi_i^0 \tilde \chi_j^0 })
- 2 G_{0ER}^{\ast ihl}G_{0ER}^{ jhl} \sqrt{x_{\tilde \chi_i^0 \tilde \chi_j^0}}
F^{\prime}(x_{\tilde{d}_k \tilde \chi_j^0 }, x_{\tilde l_h
\tilde \chi_j^0}, x_{\tilde \chi_i^0 \tilde \chi_j^0 }) \right ].
\end{eqnarray}
The right handed supersymmetric contribution is
\begin{eqnarray}
A_{{\tilde \chi}^-}^{R \Box} &=&\frac{ \alpha_W^2}{4}\sum_{h,k=1}^6  \sum_{i,
j=1}^5 \frac {1}{m_{\tilde \chi_i^-}^2} (G_{UR}^{jkb}-H_{UL}^{jkb})(G_{UR}^{\ast
iks}-H_{UL}^{\ast iks}) \nonumber \\
& & \times G_{NR}^{\ast jhl} G_{NR}^{ihl}
    G^{\prime}(x_{\tilde{u}_k \tilde \chi_j^- }, x_{\tilde{\nu}_h \tilde
\chi_j^-}, x_{\tilde \chi_i^- \tilde \chi_j^- }),  \\
A_{{\tilde \chi}^0}^{R \Box} &=& \frac{\alpha_W^2}{2} \sum_{i,j=1}^9
\sum_{k,h=1}^6
\frac {1}{m_{\tilde \chi_j^0}^2}
(\sqrt{2}G_{0DR}^{jkb}-H_{0DL}^{jkb})(\sqrt{2}G_{0DR}^{\ast iks}-H_{0DL}^{\ast
iks}) \left [G_{0ER}^{\ast jhl}G_{0ER}^{ ihl} \right. \nonumber \\
&&\left. \times  G^{\prime}(x_{\tilde{d}_k \tilde \chi_j^0}, x_{\tilde l_h
\tilde \chi_j^0}, x_{\tilde \chi_i^0 \tilde \chi_j^0 })
- 2 G_{0ER}^{\ast ihl}G_{0ER}^{ jhl}  \sqrt{x_{\tilde \chi_i^0 \tilde \chi_j^0}}
F^{\prime}(x_{\tilde{d}_k \tilde \chi_j^0 }, x_{\tilde l_h
\tilde \chi_j^0}, x_{\tilde \chi_i^0 \tilde \chi_j^0 }) \right ], \\
A_{{\tilde \chi}^0}^{R \Box \prime} &=& -\frac{\alpha_W^2}{2} \sum_{i,j=1}^9
\sum_{k,h=1}^6
\frac {1}{m_{\tilde \chi_j^0}^2}
(\sqrt{2}G_{0DR}^{jkb}-H_{0DL}^{jkb})(\sqrt{2}G_{0DR}^{\ast iks}-H_{0DL}^{\ast
iks})\left [G_{0EL}^{\ast jhl}G_{0EL}^{ ihl} \right. \nonumber \\
&&\left. \times  G^{\prime}(x_{\tilde{d}_k \tilde \chi_j^0 }, x_{\tilde l_h
\tilde \chi_j^0}, x_{\tilde \chi_i^0 \tilde \chi_j^0 })
-2 G_{0EL}^{\ast ihl}G_{0EL}^{ jhl}  \sqrt{x_{\tilde \chi_i^0 \tilde \chi_j^0}}
F^{\prime}(x_{\tilde{d}_k \tilde \chi_j^0 }, x_{\tilde l_h
\tilde \chi_j^0}, x_{\tilde \chi_i^0 \tilde \chi_j^0 }) \right ].
\end{eqnarray}
All the relevant vertex and loop functions are listed in the Appendix.

\subsection{Branching ratios and asymmetries}

Putting all the above contributions together, we write the total amplitude
at the $M_W$ scale as
\begin{eqnarray}
{\cal A}_{tot}(M_W)&=& C_{7}(M_W){Q}_{7}+ {\tilde C}_7 (M_W){\tilde Q}_{7}
+C_{9}(M_W) {Q}_{9} +{\tilde C}_{9}(M_W){\tilde Q}_{9}\nonumber \\
&+& C_{10}(M_W) {Q}_{10}+{\tilde C}_{10}(M_W){\tilde Q}_{10},
\end{eqnarray}
where
\begin{eqnarray}
{\cal A}_{tot}(M_W)&=& C_{7}(M_W){Q}_{7}+ {\tilde C}_7 (M_W){\tilde Q}_{7}
+C_{9}(M_W) {Q}_{9} +{\tilde C}_{9}(M_W){\tilde Q}_{9}\nonumber \\
&+& C_{10}(M_W) {Q}_{10}+{\tilde C}_{10}(M_W){\tilde Q}_{10},
\end{eqnarray}
where
\begin{eqnarray}
{C}_{7}(M_W)&=&A_{LR}^{\gamma}, \\
{\tilde C}_{7}(M_W)&=&A_{RL}^{\gamma}, \\
{C}_{9}(M_W)&=&A_{LL}^{\gamma}+ (-\frac14+\sin^2\theta_W)
A^{Z_L}+ \frac12 \left (A^{\Box}_{SM}+A^{L
\Box}_{\tilde \chi^-} +A^{L \Box}_{\tilde \chi^0} +A^{L \Box
\prime}_{\tilde \chi^0}\right ),\\
{\tilde C}_{9}(M_W)&=&A_{RR}^{\gamma}+ \frac12 (1+\sin^2 \theta_W ) A^{Z_R}+
\frac12 \left ( A^{R \Box}_{\tilde \chi^-}+A^{R \Box}_{\tilde \chi^0} +A^{R\Box
\prime}_{\tilde \chi^0} \right ), \\
{C}_{10}(M_W)&=& \frac12 A^{Z_L}- \frac12 \left (A^{\Box}_{SM}+ A^{L
\Box}_{\tilde \chi^-} +A^{L \Box}_{\tilde \chi^0} -A^{L \Box
\prime}_{\tilde \chi^0} \right ),\\
{\tilde C}_{10}(M_W)&=& \sin^2 \theta_W A^{Z_R}+\frac12 \left ( A^{R
\Box}_{\tilde \chi^-}+A^{R \Box}_{\tilde \chi^0} -A^{R \Box
\prime}_{\tilde \chi^0} \right ),
\end{eqnarray}
with
\begin{eqnarray}
A_{LL}^{\gamma}&=&A^{LL}_{SM}+A^{LL}_{H^-}+A^{LL}_{\tilde
g}+A^{LL}_{\tilde \chi^-}
+A^{LL}_{\tilde \chi^0},\\
A_{RR}^{\gamma}&=&A^{RR}_{\tilde g}+A^{RR}_{\tilde \chi^-}
+A^{RR}_{\tilde \chi^0},\\
A_{LR}^{\gamma}&=&A^{LR}_{SM}+A^{LR}_{H^-}+A^{LR}_{\tilde g}
+A^{LR}_{\tilde \chi^-}
+A^{LR}_{\tilde \chi^0},\\
A_{RL}^{\gamma}&=&A^{RL}_{\tilde g}+A^{RL}_{\tilde \chi^-}
+A^{RL}_{\tilde \chi^0},\\
A^{Z_L}&=&A_{SM}^{Z_L}+A_{H^-}^{Z_L}+A_{\tilde g}^{Z_L}+A_{\tilde
\chi^-}^{Z_L}+A_{\tilde \chi^0}^{Z_L},\\
A^{Z_R}&=&A_{\tilde g}^{Z_R}+A_{\tilde
\chi^-}^{Z_R}+A_{\tilde \chi^0}^{Z_R}.
\end{eqnarray}
As the experimental results on semileptonic decays $B \rightarrow X_s l^+ l^-$
fit the SM well, new physics effects can be parameterized by $R_i$
and ${\tilde R}_i$
defined at the electroweak scale as
\begin{equation}
R_i=\frac{C_i-C_i^{SM}}{C_i^{SM}}, \hspace{2cm} {\tilde
R}_i=\frac{{\tilde C}_i}{C_i^{SM}}.
\end{equation}
Note that there are no contributions to $\tilde C_i$ in the SM. The non-resonant
branching ratios can be expressed
in terms of the parameterization as \cite{GK}
\begin{eqnarray}
\mathrm{BR} (B \rightarrow X_s e^+ e^-) &=& 7.29 \times 10^{-6}
[1+0.35 R_7+0.179 R_9+0.714 R_{10}
\nonumber \\
& & + 0.0947 (R_7^2+{\tilde R}_7^2)+0.045 (R_9^2+{\tilde R}_9^2)+0.357
(R_{10}^2+{\tilde R}_{10}^2)
\nonumber \\
& & - 0.0313 (R_7 R_9 + {\tilde R}_7 {\tilde R}_9) ], \\
\mathrm{BR} (B \rightarrow X_s {\mu}^+ {\mu}^-) &=& 4.89 \times
10^{-6} [1+0.0982 R_7+0.264 R_9
+1.07 R_{10}
\nonumber \\
& & + 0.0491 (R_7^2+{\tilde R}_7^2)+0.0671 (R_9^2+{\tilde
R}_9^2)+0.535 (R_{10}^2+{\tilde R}_{10}^2)
\nonumber \\
& & - 0.0467 (R_7 R_9 + {\tilde R}_7 {\tilde R}_9) ].
\end{eqnarray}
If $R_i$ and ${\tilde R}_i$ are set to zero, the SM values for the
semileptonic decays are recovered
in these formulas. Resonant contributions were studied in Ref.
\cite{resonant} and these can be
avoided by excluding some special areas from the phase
integration regions in the dilepton
invariant mass.

We also consider the lepton pair energy asymmetry in the decay $B
\rightarrow X_s l^+ l^-$ defined as
\begin{equation}
{\cal A }_{l^+ l^-}=\frac{N(E_{l^-} > E_{l^+})-N(E_{l^+} > E_{l^-})}
{N(E_{l^-} > E_{l^+})+N(E_{l^+} > E_{l^-})},
\end{equation}
where, for instance, $N(E_{l^-} > E_{l^+})$ is the number of the
lepton pairs where the negative charged
lepton is more energetic than the positive charged lepton in the B rest frame.
The energy asymmetry is equivalent to the
ordinary forward-backward asymmetry. In a configuration where $l^+$
is scattered in the forward
direction, kinematically, in the dilepton center-of-mass frame, it is
implied that $E_{l^+} < E_{l^-}$
in the B rest frame. With the above parameterization, the energy
asymmetry is found to be
\begin{eqnarray}
{\cal A}_{l^+ l^-}&=&\frac{0.48 \times
10^{-6}}{\mathrm{R}_{\mathrm{BR}} ( B \rightarrow X_s l^+ l^-)}
[1-0.625 R_7+0.884 R_9+0.911 R_{10} \nonumber \\
& & -0.625 (R_7 R_{10}+{\tilde R}_7 {\tilde R}_{10}) +0.884 (R_9
R_{10}+{\tilde R}_9 {\tilde R}_{10})
\nonumber \\
& & -0.00882 (R_{10}^2+{\tilde R}_{10}^2)],
\end{eqnarray}
where $\mathrm{R}_{\mathrm{BR}} (B \rightarrow X_s l^+
l^-)=\frac{\mathrm{BR}(B \rightarrow X_s l^+ l^-)}
{\mathrm{BR}(B \rightarrow X_s l^+ l^-)^{SM}}$.

\section{Numerical Analysis}

We are interested in analyzing the case in which the supersymmetric
partners have masses around the weak scale, so we will assume relatively light
superpartner masses. We diagonalize the neutralino, chargino, scalar quark and lepton
mass matrices numerically and require in all calculations that the masses of
gluinos, charginos,
neutralinos, squarks and sleptons be above their experimental bounds. There are
some extra constraints in the
non-supersymmetric sector of the theory, requiring the FCNC Higgs
boson $\Phi_d$ to be heavy,
but no such constraints exist in the Higgsino sector \cite{pospelov}.
We constrain the lightest Higgs boson mass to be $115$ GeV \cite{higgs}.

As the first step, we assume the only source of flavor violation to
come from the CKM matrix. This scenario is related to the minimal
flavor violation case in supergravity. This restricted possibility
of flavor violation will set important constraints
on the parameter space of the LRSUSY model.

We then allow, in the second stage of our investigation, for new
sources of flavor violation, coming from the
soft breaking terms. In the MSSM, this scenario is known as the
unconstrained MSSM.
We restrict all allowable LL, LR, RL and RR flavor
mixings, assuming them to
be dominated by mixings between the second and third squark family.

We now proceed to discuss both these scenarios in turn.

\subsection{The constrained LRSUSY model}

By the constrained LRSUSY model, we mean the scenario
in which the only source of flavor violation comes from the quark sector,
through the CKM matrix, which we assume
to be the same for both the left- and right-handed sectors (manifest left-right symmetry), as explained below.

Before any meaningful numerical results be obtained, explicit values for
the parameters in the model must be specified. There are many parameters in
the model, such that it is hard, if not impossible, to get an illustrative
presentation of calculation results. If the LRSUSY model is embedded in a supersymmetric
grand unification theory such as $SO(\mathrm{10})$,  there exist some relationships
among the parameters at the unification scale $M_{GUT}$. We can generally choose
specific values for
parameters at the mass scale $\mu= M_{GUT}$, then use renormalization
group equations to run them down to the low energy scale
which is relevant to phenomenology. But, for maintaining both simplicity and
generality, we can present an
analysis in which the LRSUSY model is not embedded into another group. Then we
can choose all parameters as independently free variables, with the numerical
results confronting experimental bounds directly.

To make the results tractable, we assume all trilinear scalar couplings
in the soft supersymmetry breaking Lagrangian as $A_{ij}=A \delta_{ij}$ and
$\mu_{ij}=\mu \delta_{ij}$,
We also set a common mass parameter for all the squarks
$M_{0UL}=M_{0UR}=M_{0DL}=M_{0DR}=m_{0q}$.
We take $K^L_{CKM}=K^R_{CKM}$. This choice is conservative, and
much larger values
of mixing matrix elements are allowed in scenarios that attempt to
explain the decay
properties of the $b$ quark as being saturated by the right-handed $b$
\cite{gronau}. Our choice does not favor one handedness over the other,
and has the added advantage that no new mixing angles are introduced in
the quark matrices.

\begin{figure}
\centerline{ \epsfysize 4.0in \rotatebox{270}{\epsfbox{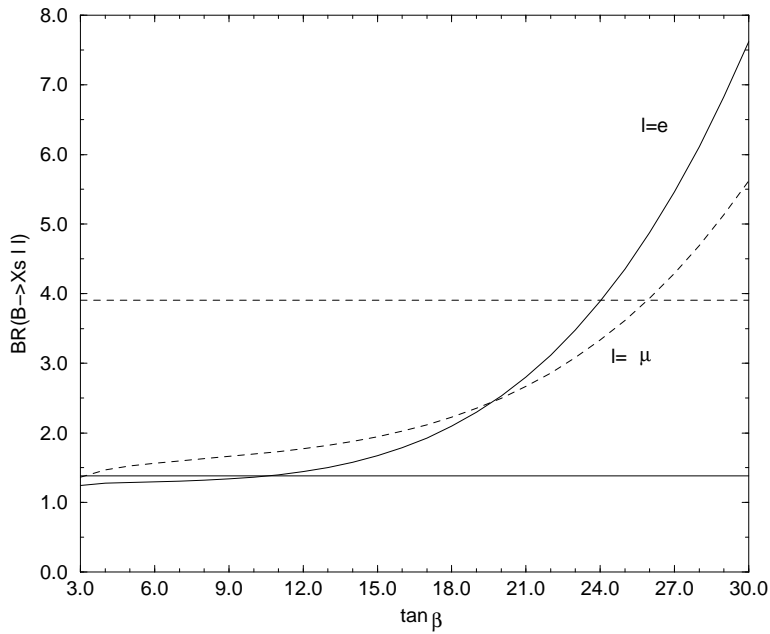}}  }
\caption{BR($B \rightarrow X_s l^+ l^-$) normalized to the corresponding SM values
as a function of $\tan \beta$,
obtained when $m_{\tilde{g}}=300$ GeV,
$\mu=100$ GeV, A=50 GeV,  $M_L=M_R=500$ GeV, $m_{0q}=300$ GeV and $m_{0l}=100$ GeV.
The experimental constraints are also shown.}
\protect \label{figbeta}
\end{figure}

\begin{figure}
\centerline{ \epsfysize 4.0in \rotatebox{270}{\epsfbox{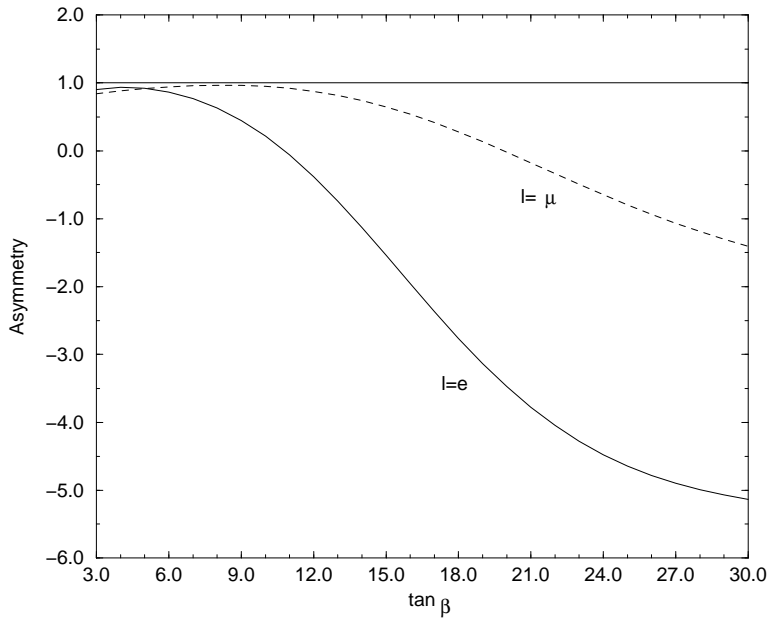}}  }
\caption{Energy asymmetry A($B \rightarrow X_s l^+ l^-$)
normalized to the corresponding SM values
as a function of $\tan \beta$,
obtained when $m_{\tilde{g}}=300$ GeV,
$\mu=100$ GeV, A=50 GeV,  $M_L=M_R=500$ GeV, $m_{0q}=300$ GeV and $m_{0l}=100$ GeV.  }
\protect \label{figAsybeta}
\end{figure}

We investigate first the dependence of the branching ratio on the
values of $\tan \beta$ in Fig. \ref{figbeta}. The braching ratios are normalized
to the corresponding SM values. In the whole parameter range, the branching ratio is greater than
one, which means that large enhancements can be obtained in the LRSUSY model with respect the SM. This feature is similar to the one in the MSSM and is due to the $1/ \cos \beta$ enhancement in the chargino interaction vertices.
For example, when $\tan \beta$ is around $30$, an enhancement of one order of magnitude can be obtained.
This would make the rare semileptonic decay more easily to be observed in future experiments.
Generally the branching ratio increases with $\tan \beta$, and for larger values of $\tan \beta$ the
branching ratio will exceed the acceptable range easily.
The choice of parameters puts
stringent restrictions on the allowed values for $\tan
\beta$. For the electron, $\tan \beta$ should be less than $11$ if the branching ratio is to be
below the experimental bounds, while for the muon $\tan \beta$ should be less than $26$, mainly due to the experimental bound of the muon is larger than that of the electron.
The asymmetries corresponding to different values of $\tan \beta$ are shown in Fig. \ref{figAsybeta}. A
clear deviation from the SM model is also obtained. Note that in the SM the asymmetry is normalized to $1$. The asymmetries tend to be large and negative with increasing $\tan \beta$.

\begin{figure}
\centerline{ \epsfysize 4.0in \rotatebox{270}{\epsfbox{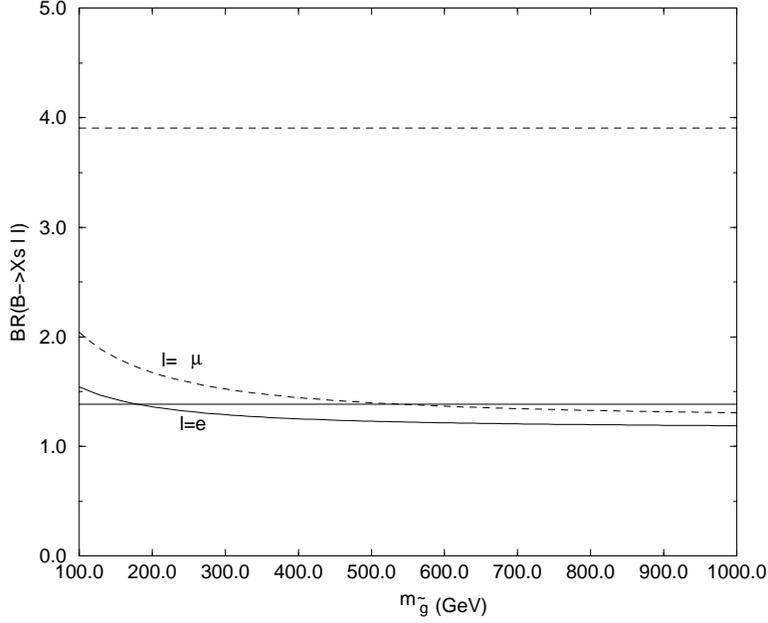}}  }
\caption{BR($B \rightarrow X_s l^+ l^-$) normalized to the corresponding SM values
as a function of $m_{\tilde g}$,
obtained when $\tan \beta=5$,
$\mu=100$ GeV, A=50 GeV, $M_L=M_R=500$ GeV, $m_{0q}=300$ GeV and $m_{0l}=100$ GeV.
The experimental constraints are also shown.}
\protect \label{figmgluino}
\end{figure}

\begin{figure}
\centerline{ \epsfysize 4.0in \rotatebox{270}{\epsfbox{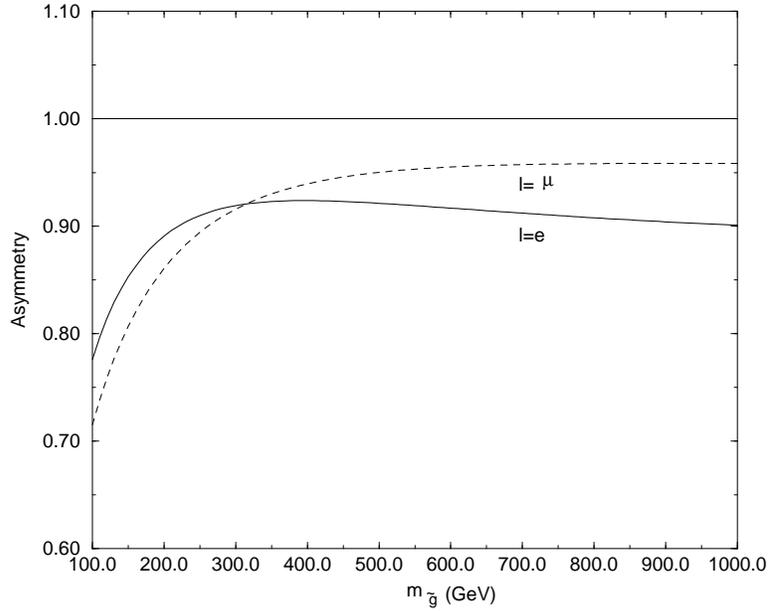}}  }
\caption{Energy asymmetry A($B \rightarrow X_s l^+ l^-$)
normalized to the corresponding SM values
as a function of $m_{\tilde g}$,
obtained when $\tan \beta=5$,
$\mu=100$ GeV, A=50 GeV, $M_L=M_R=500$ GeV, $m_{0q}=300$ GeV and $m_{0l}=100$ GeV. }
\protect \label{figAsymgluino}
\end{figure}

We investigate next the dependence of
the branching ratio and asymmetry on the gluino mass,
for a light squark scenario. The chargino and
neutralino masses are light and $\mu >0$, a scenario
favored by recent analyses of the
anomalous magnetic moment of the muon \cite{g-2} and consistent with $b \rightarrow s \gamma$ \cite{fn}. We present the
results in Fig. \ref{figmgluino}. As the mass of gluino increases, the branching ratio will
decrease, as the gluino is exchanged as a virtual particle in the process.
From the branching ratio into electrons, the gluino mass
is constrained to be heavier than 200 GeV, which is weaker than other constraints, for
example, from $b \rightarrow s \gamma$; while from the muon case there is no constraint on
$m_{\tilde g}$. Therefore in the LRSUSY model the contributions from gluino-exchanged graphs
are not dominant, while this is generally so in $b \rightarrow s \gamma$.
The corresponding asymmetries are shown in Fig. \ref{figAsymgluino}. It is found that the asymmetries for
both the electron and muon are less than the corresponding SM value. Although asymmetries do not help
the experimentalists  to observe the decay,  they might, if observed, serve to
distinguish the LRSUSY model from the SM.

\begin{figure}
\centerline{ \epsfysize 4.0in \rotatebox{270}{\epsfbox{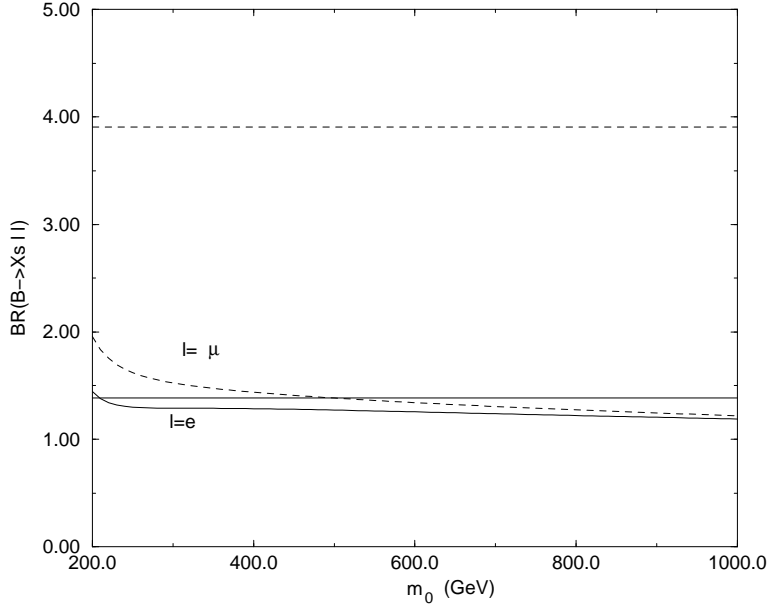}}  }
\caption{BR($B \rightarrow X_s l^+ l^-$) normalized to the corresponding SM values
as a function of $m_0$,
obtained when $m_{\tilde{g}}=300$ GeV, $\tan \beta=5$,
$\mu=100$ GeV, A=50 GeV, $M_L=M_R=500$ GeV and $m_{0l}=100$ GeV.
The experimental constraints are also shown.}
\protect \label{figm0}
\end{figure}

\begin{figure}
\centerline{ \epsfysize 4.0in \rotatebox{270}{\epsfbox{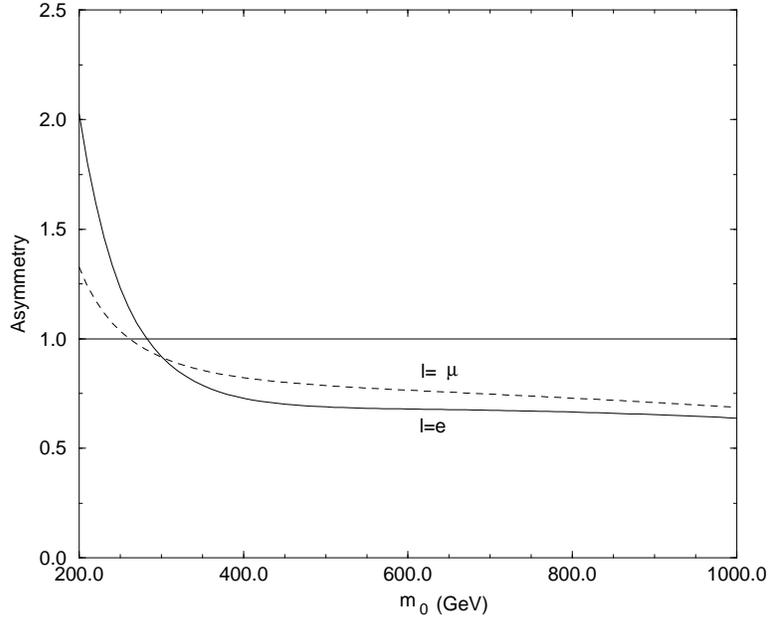}}  }
\caption{Energy asymmetry A($B \rightarrow X_s l^+ l^-$)
normalized to the corresponding SM values
as a function of $m_0$, $\tan \beta=5$,
obtained when $m_{\tilde{g}}=300$ GeV,
$\mu=100$ GeV, A=50 GeV, $M_L=M_R=500$ GeV and  $m_{0l}=100$ GeV.  }
\protect \label{figAsym0}
\end{figure}

The branching ratio of $B \rightarrow X_s l^+ l^-$ is sensitive to the
universal scalar mass $m_{0q}$ in the region of small masses only, a feature shared with $b \rightarrow s
\gamma$ \cite{fn}.  This dependence is shown in Fig. \ref{figm0}. For the electron case, $m_{0q}$ is
found to be greater than $200$ GeV, where the corresponding scalar quark masses are slightly
above the current experimental bounds, while for muon case there is no constraint on $m_{0q}$.
The lepton asymmetries as a function of the universal scalar mass, are shown in Fig. \ref{figAsym0}, and
there a small enhancement can be found when
$m_{0q}$ is less than $300$ GeV.

\begin{figure}
\centerline{ \epsfysize 4.0in \rotatebox{270}{\epsfbox{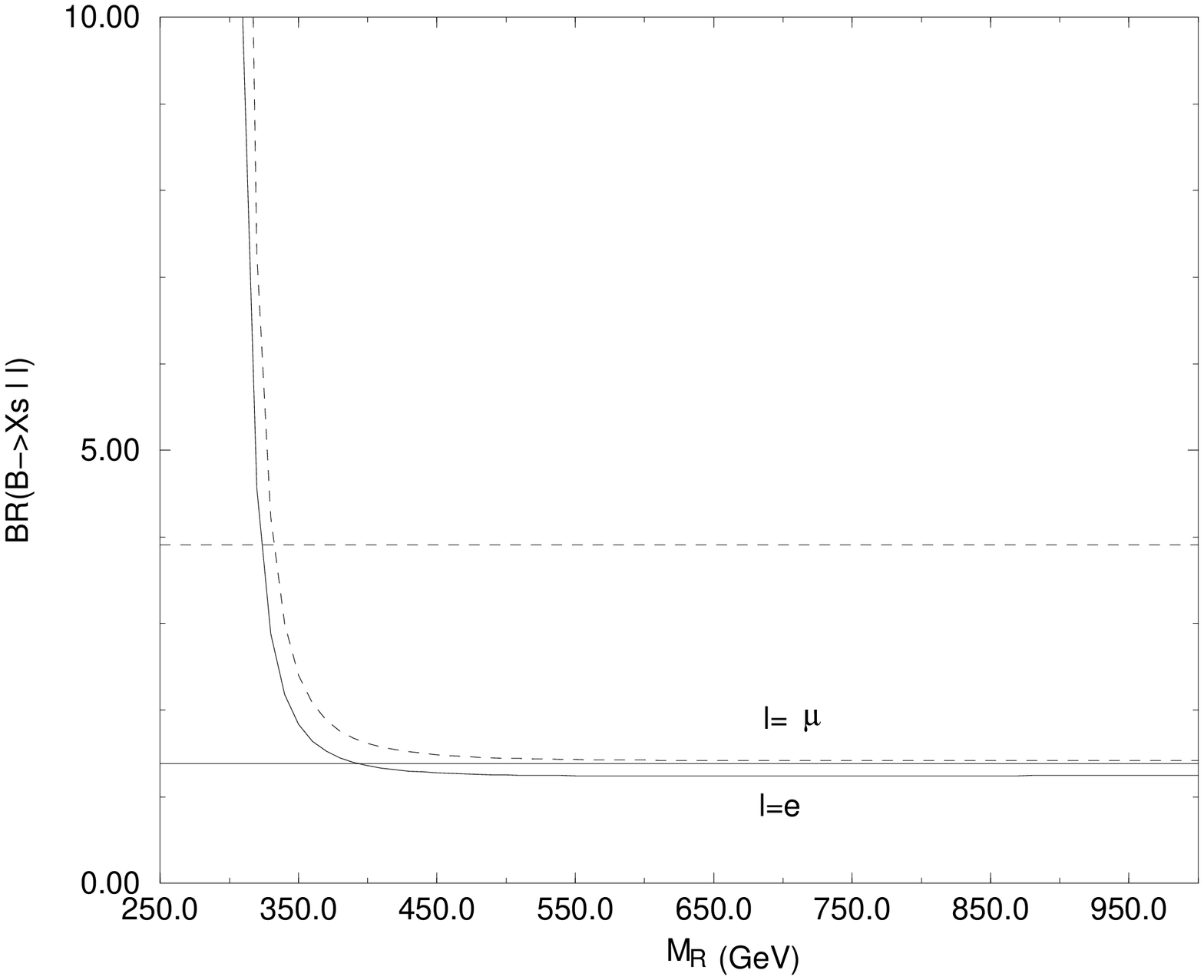}}  }
\caption{BR($B \rightarrow X_s l^+ l^-$) normalized to the corresponding SM values
as a function of $M_R$,
obtained when $m_{\tilde{g}}=400$ GeV, $\tan \beta=5$,
$\mu=100$ GeV, A=50 GeV, $m_{0q}=300$ GeV and $m_{0l}=100$ GeV.
The experimental constraints are also shown.}
\protect \label{figMR}
\end{figure}

\begin{figure}
\centerline{ \epsfysize 4.0in \rotatebox{270}{\epsfbox{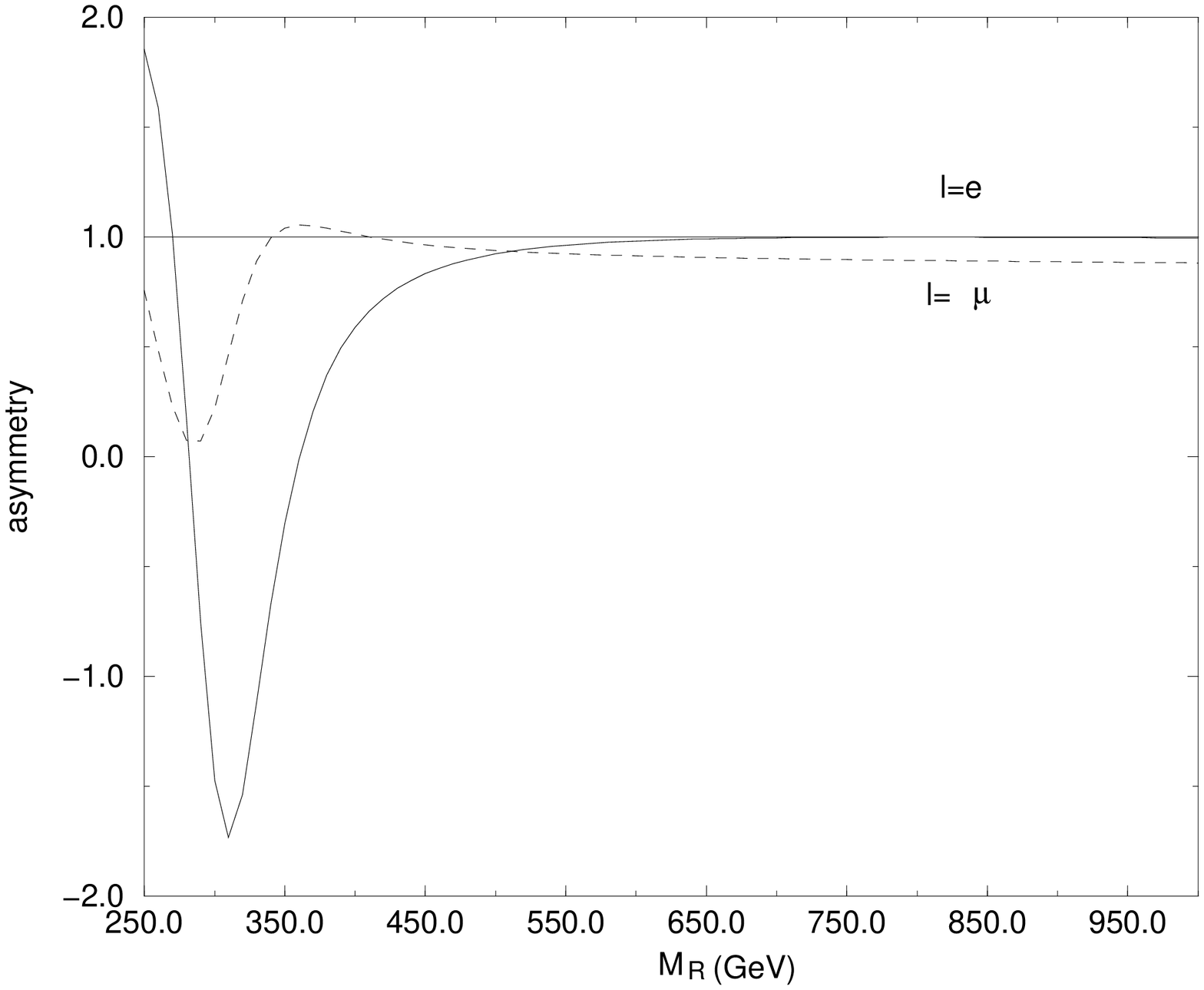}}  }
\caption{Energy asymmetry A($B \rightarrow X_s l^+ l^-$)
normalized to the corresponding SM values
as a function of $M_R$,
obtained when $m_{\tilde{g}}=400$ GeV, $\tan \beta=5$,
$\mu=100$ GeV, A=50 GeV, $m_{0q}=300$ GeV and $m_{0l}=100$ GeV. }
\protect \label{figAsyMR}
\end{figure}

In all the previous figures we set the left- and right-handed gaugino
masses to the same value.  In
Fig. \ref{figMR} and Fig. \ref{figAsyMR} we investigate
the dependence of the branching ratios and asymmetries on the
gaugino mass. When $M_R \cong m_{0q}$, the results are not reliable, due to poles in the loop functions. As the branching ratio of $B \rightarrow X_s l^+ l^-$ is dominated by the chargino contribution for a large region of the parameter space, one expects a restriction on the left- and right-handed gaugino mass parameters.  
There exist scenarios in which the right-handed symmetry is
broken at the same scale as supersymmetry, so
we expect in those cases to have approximately $M_L=M_R$ \cite{kai}.
With the assumption $M_L \cong M_R$ in the gaugino sector, the restriction on the 
right-handed gaugino scale is found to be $M_R > 400-500$ GeV, for low to intermediate squark masses.

\begin{figure}
\centerline{ \epsfysize 4.0in
\rotatebox{270}{\epsfbox{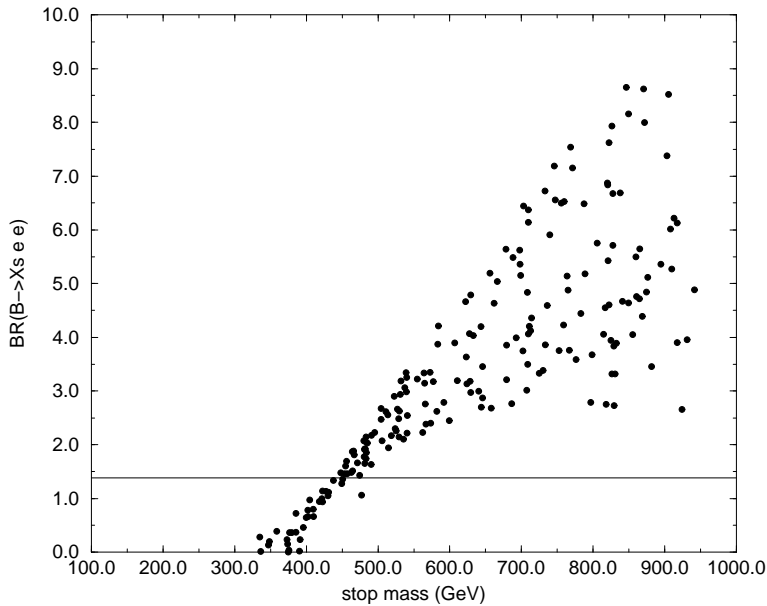}}  }
\caption{BR($B \rightarrow X_s e^+ e^-$) normalized to the corresponding SM values
versus the lightest stop mass in the LRSUSY model. }
\protect \label{figBRvSquark}
\end{figure}

\begin{figure}
\centerline{ \epsfysize 4.0in
\rotatebox{270}{\epsfbox{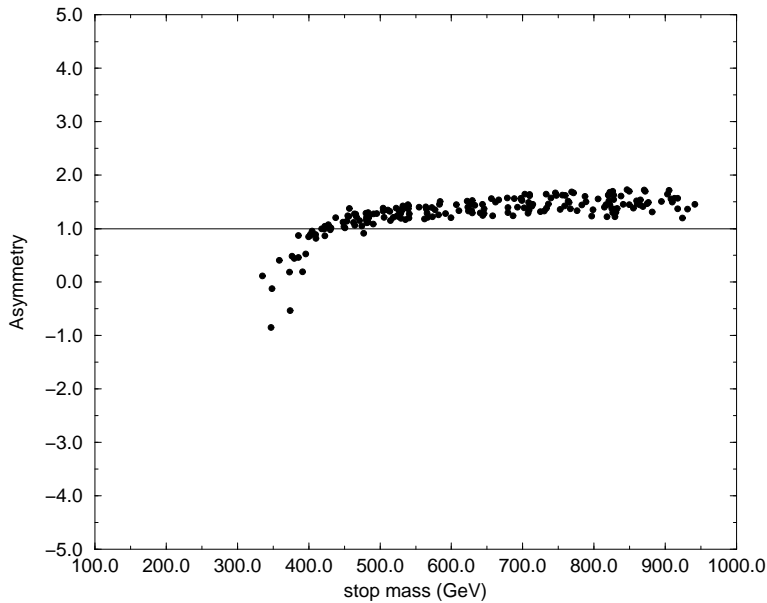}}  }
\caption{Energy asymmetry A($B \rightarrow X_s e^+ e^-$) normalized to the corresponding SM values
versus the lightest stop mass in the LRSUSY model. }
\protect \label{AsyvSquark}
\end{figure}

In Fig. \ref{figBRvSquark} and Fig. \ref{AsyvSquark} we present the scatter plots of the branching ratio and the
asymmetry for the decay $B \rightarrow X_s e^+ e^-$ as a function of the lightest stop mass.
We have chosen randomly relevant parameters: $\tan \beta$ changes from $2$ to $30$, $m_{0q}$ takes values
from $100$ to $1000$ GeV, $\mu$ also varies from $100$ to $1000$ GeV and $A=m_{0q}$, while
$m_{\tilde g}=500$ GeV, $M_L=M_R=500$ GeV and $m_{0l}=100$ GeV. In addition to the
current experimental bounds on the SUSY spectra, we also impose the constraints from the rare decay $b
\rightarrow s \gamma$.  It is found that the lightest stop masses lower than $300$ GeV are excluded. The
branching ratio fits easily within the experimental bound for large $m_{\tilde t}$, which explains the
large number of plot points in that region. An enhancement for the branching ratio of one order of magnitude is possible
while the asymmetry could be $50 \%$ larger than the SM value.

\begin{figure}
\centerline{ \epsfysize 4.0in
\rotatebox{270}{\epsfbox{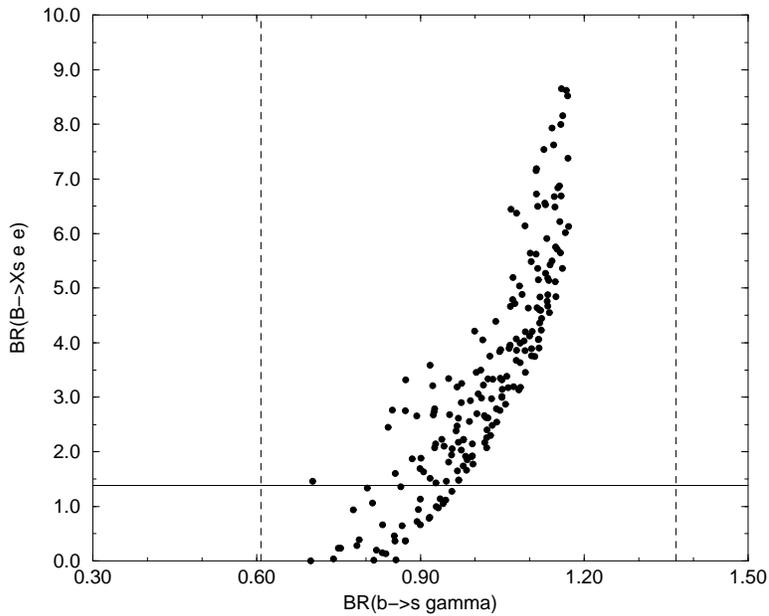}}  }
\caption{A correlation between BR($b \rightarrow s \gamma$) and
BR($B \rightarrow X_s e^+ e^-$) in the LRSUSY model. Experimental bounds are also shown. }
\protect \label{correlation}
\end{figure}

In Fig. \ref{correlation} we show the correlation between the branching ratio of the $b \rightarrow s \gamma$
and $B \rightarrow X_s e^+ e^-$ in the above specified parameter ranges. Both branching ratios are
normalized to the  corresponding SM values. Although it seems possible that $b \rightarrow s \gamma$ is
below the SM value, while $B \rightarrow X_s e^+ e^-$ is enhanced with respect to the SM value, there
exists a region of the parameter space in which both are significantly
enhanced with respect to their SM values.

\subsection{The unconstrained LRSUSY model}

When supersymmetry is softly broken, there is no reason to expect
that the soft parameters would be flavor blind, or that they would
violate flavor in the same way as in the SM.
Yukawa couplings generally form a matrix in the generation space, and
the off-diagonal elements will lead naturally to flavor changing
radiative decays.
Neutrino oscillations, in particular, indicate strong flavor mixing
between the second and
third neutrino generations, and various analyses have been carried out
assuming the same for the
charged sleptons. In the quark/squark sector, the kaon system
strongly limits mixings
between the first and the second generations; constraints for the
third generation from $b \rightarrow s \gamma$ in the LRSUSY model are studied in Ref. \cite{fn}.

We parametrize all the unknown soft breaking parameters, coming mostly from the
scalar mass matrices, using the mass insertion approximation (MI) \cite{MI}.
In this framework we choose a basis for fermion and sfermion states
in which all the couplings of these particles to neutral gauginos are flavor
diagonal. Flavor
changes in the squark sector arise from the non-diagonality of the squark
propagators. These off-diagonal squark mass matrix elements are
assumed to be small and their higher orders can be neglected, and the
normalized parameters used in the analysis are:
\begin{eqnarray}
\label{massins}
\delta_{d,LL,ij}&=&\frac{(m^2_{d,LL})_{ij}}{m_{0q}^2},~~~
\delta_{d,RR,ij}=\frac{(m^2_{d,RR})_{ij}}{m_{0q}^2}, \nonumber \\
\delta_{d,LR,ij}&=&\frac{(m^2_{d,LR})_{ij}}{m_{0q}^2},~~~
\delta_{d,RL,ij}=\frac{(m^2_{d,RL})_{ij}}{m_{0q}^2},
\end{eqnarray}
where $(m^2_{d,AB})_{ij}$, $A,B=L,R$ are
the off-diagonal elements which mix down-squark flavors for both left- and
right-handed
squarks. We assume significant mixings between the second and third
generation in
the down-squarks mass matrix only. We also consider terms with one
mass insertion only. Although it was shown in the MSSM that double mass insertions
could possibly
enhance the decays of the $K$ meson, their effects on $B$ meson decays are
assumed to be negligible. This procedure allows an analysis of the graphs
contributing to $b \rightarrow s l^+l^-$ in terms of a small number of
parameters.
The contribution of each graph with the MI is obtained from the constrained
case following these simple rules:
\begin{itemize}
\item A left gaugino-gaugino vertex has a factor of $\delta_{d,LL,23}$
associated with it; a right gaugino-gaugino vertex has a factor of
$\delta_{d,RR,23}$.
\item A left gaugino Higgsino vertex has a factor of $\delta_{d,LR,23}$
associated with it; a right gaugino-Higgsino vertex has a factor of
$\delta_{d,RL,23}$.
\item In addition, in the dipole contributions, the term coming from
chirality being
flipped on the fermion leg, proportional to $m_{\tilde \chi}/m_b$ or $m_{\tilde
g}/m_b$, has a factor $\delta_{d,LR,23}$ associated with it for the LR
contribution, and $\delta_{d,RL,23}$ associated with it for the
RL contribution.
\end{itemize}

\begin{figure}
\centerline{ \epsfysize 4.0in \rotatebox{270}{\epsfbox{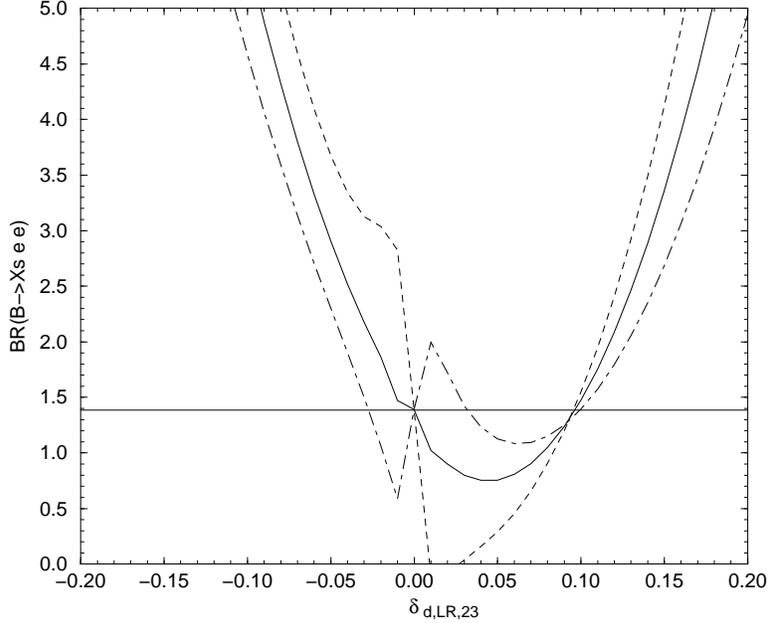}}  }
\caption{BR($B \rightarrow X_s e^+ e^-$) normalized to the corresponding SM values
as a function of $\delta_{d,LR,23}$,
obtained when $\tan \beta =5$,
$\mu=100$ GeV, A=50 GeV, $M_L=M_R=500$ GeV, $m_{0q}=500$ GeV and $m_{0l}=100$ GeV.
The different lines correspond to different values
of $x=m^2_{\tilde{g}}/m^2_{0q}=0.64(dashed), 1(solid), 1.44(dot-dashed)$.
The experimental constraints are also shown.}
\protect \label{figdeltadLR23}
\end{figure}

With the definition of the mass insertion as in Eq. (\ref{massins}),
we can investigate the
effects of intergenerational mixings on the $B \rightarrow X_s l^+ l^-$ decays.
We keep our analysis general, but to
show our results, we select only one possible source of flavor
violation in the squark sector at a time, and assume the others vanish.
In Fig. \ref{figdeltadLR23}  we show the
dependence of $B \rightarrow X_s e^+ e^-$ as a function of $\delta_{d,
LR,23}$, when this is the
only source of flavor violation. The horizontal line represents the
experimental bound on the branching ratio. The branching ratio is plotted as a
function of different
values for the mass ratio $x=m^2_{\tilde{g}}/m^2_{0q}$. Fixing $m_{0q}=500$
GeV, this corresponds to
gluino masses of $400$, $500$ and $600$ GeV respectively.
Constraints on positive and negative
values of $\delta_{d,LR,23}$ are slightly different, $\delta_{d,LR,23}$ is constrained to be positive for small mass ratios, and the absolute value of $\delta_{d, LR,23}$ is less than $10 \%$.
This flavor violating parameter can be strongly constrained from $b \rightarrow
s \gamma$ because through the
$\delta_{d,LR,23}$ term, the helicity flip needed for $b \rightarrow
s \gamma$ can be realized in the exchange particle loop.
The constraint obtained here is complimentary to that from $b \rightarrow
s \gamma$ \cite{fn, bghw,kane}. The results for $B \rightarrow X_s \mu^+ \mu^-$ are not restrictive, so we
will not show them here.

\begin{figure}
\centerline{ \epsfysize 4.0in \rotatebox{270}{\epsfbox{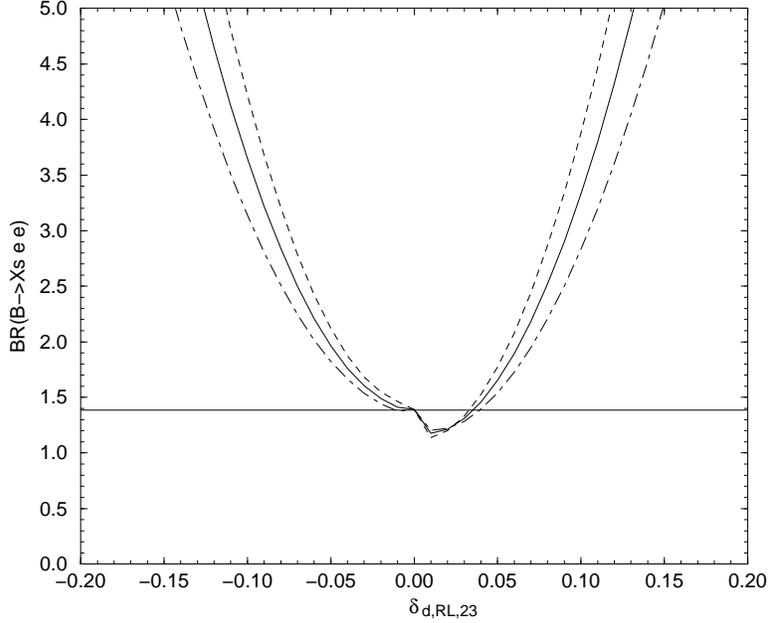}}  }
\caption{BR($B \rightarrow X_s e^+ e^-$) normalized to the corresponding SM values
as a function of $\delta_{d,LR,23}$,
obtained when $\tan \beta =5$,
$\mu=100$ GeV, A=50 GeV, $M_L=M_R=500$ GeV, $m_{0q}=500$ GeV and $m_{0l}=100$ GeV.
The different lines correspond to different values
of $x=m^2_{\tilde{g}}/m^2_{0q}=0.64(dashed), 1(solid), 1.44(dot-dashed)$.
The experimental constraints are also shown.}
\protect \label{figdeltadRL23}
\end{figure}

The situation is different when the only source of flavor violation is $\delta_{d,RL,23}$, as shown
in Fig. \ref{figdeltadRL23}. Again, the most restrictive case is for $B \rightarrow X_s e^+ e^-$ and
for the same values of squark and gluino masses as before, only positive values of $\delta_{d,RL,23}$ in a small interval close
to zero would satisfy the experimental bounds, namely $\delta_{d,RL,23} < 4 \%$.

The restrictions on the branching ratio of $B \rightarrow X_s l^+ l^-$ from the chirality conserving
mixings $\delta_{d,LL,23}$ and
$\delta_{d,RR,23}$ respectively, with the proviso that these are the
only off-diagonal matrix elements in the squark mass matrix squared, are not as
pronounced as the ones for chirality flipping parameters.
$\delta_{d,LL,23}$ and $\delta_{d,RR,23}$ can almost take all the values in the range
$(-1.0, 1.0)$. We don't show the results here.

\section{Conclusions}

We analyze the FCNC semileptonic decay $B
\rightarrow X_s l^+l^-$ in a fully left-right supersymmetric
model. Explicit expressions for all the amplitudes involved in
the process are given. Constraints on the parameter 
space of the model are obtained in
both the constrained case (where the only flavor violation comes from the
CKM matrix) and the unconstrained case
(including soft supersymmetry breaking terms). We also include and compare with
constraints from $b \rightarrow s \gamma$.

As a general feature, both $b \rightarrow s \gamma$ and $B \rightarrow X_s l^+ l^-$ exhibit similar
dependences on squark and gluino masses. From restrictions on both decays, we expect $m_{\tilde g} \ge
250-300$ GeV and $m_{\tilde q} \ge 200$ GeV. A more careful analysis of the branching ratio of
$B \rightarrow X_s l^+l^-$ reveals that, varying all other parameters in the model, the mass of the
lightest scalar top should be $\ge 300$ GeV, which is much more restrictive than the experimental bound \cite{particledata}. The parameter that most sensitively affects the branching
ratio of $B \rightarrow X_s l^+l^-$ is $\tan \beta$. The constraint from $B \rightarrow X_s l^+l^-$ is
slightly more restrictive than for $b \rightarrow s \gamma$, and the semileptonic branching ratio can 
exceed the experimental bound for $\tan \beta \ge 10-11$, for low squark and gluino masses. In all our
analysis, we keep $\mu>0$, and in regions allowed by both $(g-2)_{\mu}$ and $b \rightarrow s \gamma$.

An analysis of the correlation between the branching ratio of $B \rightarrow X_s \gamma$ and $B
\rightarrow X_s l^+ l^-$ reveals that there is a larger region of parameter space in which $B \rightarrow
X_s l^+ l^-$ is enhanced with respect to the SM value by factors of almost 10, while $B \rightarrow X_s
\gamma$ is at most 20\% larger than the SM value. We expect the enhancements to come mostly from
regions of intermediate or large $\tan \beta$.
The LRSUSY model shares the strong $\tan \beta$ dependence with the MSSM, except that here the enhancement is even more pronounced. The asymmetry also shows different features from the MSSM \cite{dov}: it does not peak when the Higgs and gauge induced flavor violation are of the same size ($\tan \beta =35$), since the contributions from the gaugino sector are different. Also, for low and intermediate values of supersymmetric masses ($m_0, m_{\tilde{g}}, M_L, M_R=200-500$ GeV), this value of $\tan \beta$ is ruled out by constraints from $B \rightarrow X_s e^+ e^-$.    

In the unconstrained model, allowing for flavor-dependent soft mixing between the second and third
generation of squarks (both chirality conserving and chiralty violating), no reliable limits are set on either
the LL or the RR mixing. However, the chirality violating soft mixing parameters are strongly
constrained. In particular, the RL mixing, $\delta_{d,RL,23}$ is constrained within four percentage to be close to zero from
the bounds on $B \rightarrow X_s e^+ e^-$, for a variety of squark and gluino masses; while the
constraints on $\delta_{d,LR,23}$ favor positive values up to $10 \%$.
It could be difficult to compare these values with the unconstrained MSSM \cite{lmss}. The bound on $\delta_{d,LR,23}$ appears to be much stronger in the unconstrained LRSUSY model than in the MSSM, where maximum enhancements are obtained at values of the left-right splitting ruled out in the LRSUSY model. And certainly the restriction on the $\delta_{d,RL,23}$ coming from $B \rightarrow X_s e ^+ e^-$ is, to our knowledge, new. 

It appears likely that the most distinguishing factor of the LRSUSY model from the SM would be the
forward-backward lepton asymmetry. This asymmetry, like the branching ratio, is most sensitive to
variations in $\tan \beta$ and could be spectacular even in regions of $\tan \beta$ allowed by
constraints on the branching ratio. The asymmetry tends to be large and negative with increasing $\tan
\beta$, whereas it is small and positive when varying other parameters. As always, the regions of
interest are regions of small to intermediate values (allowed for branching ratios) for gluino, chargino and squark masses.
These enhancements are much more pronounced in the LRSUSY model than in the MSSM, and increases in the asymmetry by a factor of $2$ with respect to the SM value are allowed, for a large region of the parameter space.
 
In conclusion, the decay $B \rightarrow X_s l^+ l^-$ would provide an interesting, and complementary to $b \rightarrow s
\gamma $, test of the LRSUSY model.

\vskip0.5in

\noindent {\bf Acknowledgements}

This work was funded by NSERC of Canada (SAP0105354).

\newpage
\begin{appendix}
\noindent {\Large {\bf Appendix}}

The relevant Feynman rules and loop functions used in the calculation
are listed in this Appendix. For further details, we refer to \cite{fn}.
The terms relevant to the masses of charginos in the Lagrangian are
\begin{equation}
{\cal L}_C=-\frac{1}{2}(\psi^+, \psi^-) \left ( \begin{array}{cc}
                                                       0 & X^T \\
                                                       X & 0
                                                     \end{array}
                                             \right ) \left (
\begin{array}{c}
                                                              \psi^+ \\
                                                              \psi^-
                                                              \end{array}
                                                       \right ) + H.c. \ ,
\end{equation}
where $\psi^+=(-i \lambda^+_L, -i \lambda^+_R, \tilde{\phi}_{u1}^+,
\tilde{\phi}_{d1}^+, \tilde{\Delta}_R^+)^T$
and $\psi^-=(-i \lambda^-_L, -i \lambda^-_R, \tilde{\phi}_{u2}^-,
\tilde{\phi}_{d2}^-, \tilde{\delta}_R^-)^T$, and
\begin{equation}
X=\left( \begin{array}{ccccc}
                           M_L & 0 & g_L \kappa_u & 0 & 0 \\
                           0 & M_R & g_R \kappa_u & 0 & 0 \\
                           0 & 0 & 0 & -\mu & 0 \\
                           g_L \kappa_d & g_R \kappa_d & -\mu & 0 & 0 \\
                           0 & \sqrt{2} g_R v_R & 0 & 0 & -\mu
              \end{array}
        \right )
\end{equation}
where we have taken, for simplification, $\mu_{ij}=\mu$. The chargino mass
eigenstates $\chi_i$ are obtained by
\begin{eqnarray}
\chi_i^+=V_{ij}\psi_j^+, \ \chi_i^-=U_{ij}\psi_j^-, \ i,j=1, \ldots 5,
\end{eqnarray}
with $V$ and $U$ unitary matrices satisfying
\begin{equation}
U^* X V^{-1} = M_D,
\end{equation}
The diagonalizing matrices $U^*$ and $V$ are obtained by
computing the eigenvectors corresponding
to the eigenvalues of $X^{\dagger} X$ and $X X^{\dagger}$, respectively.

The terms relevant to the masses of neutralinos in the Lagrangian are
\begin{equation}
{\cal L}_N=-\frac{1}{2} {\psi^0}^T Y \psi^0  + H.c. \ ,
\end{equation}
where $\psi^0=(-i \lambda_L^3, -i \lambda_R^3, -i \lambda_V,
\tilde{\phi}_{u1}^0, \tilde{\phi}^0_{u2},
\tilde{\phi}_{d1}^0, \tilde{\phi}^0_{d2}, \tilde{\Delta}_R^0,
\tilde{\delta}_R^0 )^T $,
and
\begin{equation}
Y=\left( \begin{array}{ccccccccc}
              M_L & 0 & 0 & \frac{g_L \kappa_u}{\sqrt{2}} & 0 & 0 & -
\frac{g_L \kappa_d}{\sqrt{2}} & 0 & 0 \\
              0 & M_R & 0 & \frac{g_R \kappa_u}{\sqrt{2}} & 0 & 0 &
-\frac{g_R \kappa_d}{\sqrt{2}} & -\sqrt{2}g_R v_R & 0 \\
              0 & 0 & M_V & 0 & 0 & 0 & 0 & 2 \sqrt{2} g_V v_R & 0 \\
              \frac{g_L \kappa_u}{\sqrt{2}} & \frac{g_R \kappa_u}{\sqrt{2}}
&
0 & 0 & 0 & 0 & -\mu & 0 & 0  \\
              0 & 0 & 0 & 0 & 0 & -\mu & 0 & 0 & 0 \\
              0 & 0 & 0 & 0 & -\mu & 0 & 0 & 0 & 0 \\
              -\frac{g_L \kappa_d}{\sqrt{2}} & -\frac{g_R
\kappa_d}{\sqrt{2}}
& 0 & -\mu & 0 & 0 & 0 & 0 & 0  \\
              0 & -\sqrt{2}g_R v_R & \sqrt{2}g_V v_R & 0 & 0 & 0 & 0 & 0 &
-\mu \\
              0 & 0 & 0 & 0 & 0 & 0 & 0 & -\mu & 0
              \end{array}
        \right ).
\end{equation}
The mass eigenstates are defined by
\begin{equation}
\chi^0_i=N_{ij} \psi^0_j \ (i,j=1,2, \ldots 9),
\end{equation}
where $N$ is a unitary matrix chosen such that
\begin{equation}
N^* Y N^{-1} = N_D,
\label{equationN}
\end{equation}
and $N_D$ is a diagonal matrix with non-negative entries.

In the interaction basis, $(\tilde{q}_L^{i}, \tilde{q}_R^{i})$, the
squared-mass matrix for a squark of flavor $f$ has the following forms.
For U-type squarks
\begin{equation}
{\cal M}_{U_k}^2= \left( \begin{array}{cc}
                               m_0^2+M_Z^2(T_u^3-Q_u \sin^2 \theta_W) \cos 2
\beta & m_{u_k} (A-\mu \cot \beta) \\
                              m_{u_k} (A-\mu \cot \beta) &
m_0^2+M_Z^2 Q_u \sin^2 \theta_W \cos 2 \beta
                          \end{array}
                   \right).
\end{equation}
and for D-type squarks
\begin{equation}
{\cal M}_{D_k}^2= \left( \begin{array}{cc}
                               m_0^2+M_Z^2(T_d^3-Q_d \sin^2 \theta_W) \cos 2
\beta & m_{d_k} (A-\mu \tan \beta)\\
                              m_{d_k} (A-\mu \tan \beta) &
m_0^2+M_Z^2 Q_d \sin^2 \theta_W \cos 2 \beta
                          \end{array}
                   \right).
\end{equation}
The corresponding mass eigenstates are defined as
\begin{equation}
{\tilde q}_{L,R}=\Gamma^{\dagger}_{Q \ L,R} \tilde{q},
\end{equation}
where $\Gamma^{\dagger}_{Q \ L,R}$ are 6$\times$ 3 mixing matrices. The same
expressions, with the switches $Q \rightarrow L$, $U \rightarrow N$ and $D
\rightarrow E$ exist for the sleptons and sneutrinos. The chargino-quark-squark
mixing martices $G$ and $H$ are defined as
\begin{eqnarray}
G^{jki}_{UL} &=& V_{j1}^{\ast} (K_{CKM})_{il} (\Gamma_{UL})_{kl},
\nonumber \\
G^{jki}_{UR} &=& U_{j2} (K_{CKM})_{il} (\Gamma_{UR})_{kl}, \nonumber \\
H_{UL}^{jki} &=& \frac{1}{\sqrt{2} m_W} ( \frac{m_{u_l}}{\sin \beta}
U_{j3}+
\frac{m_{d_l}}{\cos \beta} U_{j4} ) (K_{CKM})_{il} (\Gamma_{UL})_{kl},
\nonumber \\
H_{UR}^{jki} &=& \frac{1}{\sqrt{2} m_W} ( \frac{m_{u_l}}{\sin \beta}
V_{j3}^*+
\frac{m_{d_l}}{\cos \beta} V_{j4}^* ) (K_{CKM})_{il} (\Gamma_{UR})_{kl}.
\end{eqnarray}
and the gaugino-sneutrino-lepton $G_{NL,R}$ are defined as
\begin{eqnarray}
G^{jki}_{NL,R} &=& V_{j1}^{\ast} (\Gamma_{NL,R})_{ki}.
\end{eqnarray}
The neutralino-quark-squark mixing matrices $G_0$ and $H_0$ are defined as
\begin{eqnarray}
G^{jki}_{0DL} &=& [\sin \theta_W Q_d N^{\prime}_{j1} + \frac{1}{\cos
\theta_W} (T^3_{d}-Q_d \sin^2 \theta_W)
      N^{\prime}_{j2} \nonumber \\
& -&\frac{\sqrt{\cos 2 \theta_W }}{\cos \theta_W}
\frac{Q_u+Q_d}{2} N^{\prime}_{j3}
] (K_{CKM})_{il} (\Gamma_{DL})_{kl}, \nonumber \\
G^{jki}_{0DR} &=& -[\sin \theta_W Q_d N^{\prime}_{j1} - \frac{Q_d \sin^2
\theta_W}{\cos \theta_W}
      N^{\prime}_{j2} \nonumber \\
&+&\frac{\sqrt{\cos 2 \theta_W }}{\cos \theta_W}
(T^3_{d}-Q_d \sin^2 \theta_W) N^{\prime}_{j3}
] (K_{CKM})_{il} (\Gamma_{DR})_{kl},  \nonumber \\
H_{0DL}^{jki} &=&  \frac{1}{\sqrt{2} m_W} ( \frac{m_{u_l}}{\sin \beta}
N^{\prime}_{j5}+
\frac{m_{d_l}}{\cos \beta} N^{\prime}_{j7}) (K_{CKM})_{il}
(\Gamma_{DL})_{kl}, \nonumber \\
H_{0DR}^{jki} &=& \frac{1}{\sqrt{2} m_W} ( \frac{m_{u_l}}{\sin \beta}
N^{\prime *}_{j5}+
\frac{m_{d_l}}{\cos \beta} N^{\prime \ast}_{j7} ) (K_{CKM})_{il}
(\Gamma_{DR})_{kl}.
\end{eqnarray}
and the gaugino-slepton-lepton mixing matrices $G_{0EL},~G_{0ER}$ are defined as
\begin{eqnarray}
G^{jki}_{0EL} &=& [\sin \theta_W Q_e N^{\prime}_{j1} + \frac{1}{\cos
\theta_W} (T^3_{e}-Q_e \sin^2 \theta_W)
      N^{\prime}_{j2} \nonumber \\
& -&\frac{\sqrt{\cos 2 \theta_W }}{\cos \theta_W}
\frac{Q_e}{2} N^{\prime}_{j3}
] (\Gamma_{EL})_{ki}, \nonumber \\
G^{jki}_{0ER} &=& -[\sin \theta_W Q_e N^{\prime}_{j1} - \frac{Q_e \sin^2
\theta_W}{\cos \theta_W}
      N^{\prime}_{j2} \nonumber \\
&+&\frac{\sqrt{\cos 2 \theta_W }}{\cos \theta_W}
(T^3_{e}-Q_e \sin^2 \theta_W) N^{\prime}_{j3}
]  (\Gamma_{ER})_{ki}.
\end{eqnarray}
The one, two and three variable functions appearing in the decay $b
\rightarrow s l^+l^-$ are \cite{function}
\begin{eqnarray}
F_1(x)&=&\frac{1}{12(x-1)^4}(x^3-6x^2+3x+2+6x\log x),\\
F_2(x)&=&\frac{1}{12(x-1)^4}(2 x^3-3x^2-6x+1-6x^2\log x),\\
F_3(x)&=&\frac{1}{2(x-1)^3}(x^2-4x+3+2\log x),\\
F_4(x)&=&\frac{1}{2(x-1)^3}(x^2-1-2x\log x),\\
F_5(x)&=&\frac{1}{36(x-1)^4}[7x^3-36x^2+45x-16+(18x-12)\log x],\\
F_6(x)&=&\frac{1}{36(x-1)^4}(-11x^3+18x^2-9x+2+6x^3\log x),\\
F_7(x)&=&\frac{1}{12(x-1)^4}[x^3+10x^2-29x+18-(8x^2-6x-8)\log x],\\
F_8(x)&=&\frac{1}{12(x-1)^4}[-7x^3+8x^2+11x-12-(2x^3-20x^2+24x)\log x],\\
F_9(x)&=&\frac{1}{2(x-1)^2}[x^2-7x+6+(3x+2)\log x],\\
f(x)&=&-\frac{2}{3} -\frac{z}{x}+ \left \{ \begin{array}{cc} 2 \left
(1+\frac{z}{2x}
\right )\sqrt{\frac{z}{x}-1}\tan^{-1}\left \{\sqrt{\frac{z}{x}-1} \right
\}^{-1},&\mbox{if $x<z$}\\
\left (1+ \frac{z}{2x}\right )\sqrt{1-\frac{z}{x}}\left \{\ln
\frac{1-\sqrt{1-\frac{z}{x}}}{1-\sqrt{1-\frac{z}{x}}} -i \pi \right \},&\mbox{if $x>z$}
\end{array}
\right. \\
F_0(x,y)&=&\frac{1}{x-y}\left
[\frac{x}{x-1}\log x-(x \rightarrow y)
\right ],\\
G_0(x,y)&=&\frac{1}{x-y}\left [\frac{x^2}{x-1}\log x -
\frac32 x -(x \rightarrow y) \right ],\\
F(x,y)&=&-\frac{1}{x-y}\left [\frac{x}{(x-1)^2}\log x-
\frac{1}{x-1}-(x \rightarrow
y) \right ],\\
G(x,y)&=&\frac{1}{x-y}\left [\frac{x^2}{(x-1)^2}\log x-
\frac{1}{x-1}-(x \rightarrow
y) \right ],\\
G^{\prime}(x,y,z)&=&\frac{1}{x-y} \left \{\frac{1}{x-z}\left [ \frac{x^2}{x-1}
\log x-\frac32 x-(x \rightarrow z)\right ]- (x \rightarrow y)\right \},\\
F^{\prime}(x,y,z)&=&-\frac{1}{x-y} \left \{\frac{1}{x-z}\left [ \frac{x}{x-1}
\log x-(x \rightarrow z)\right ]- (x \rightarrow y)\right \}.
\end{eqnarray}
\end{appendix}

\newpage
\def\oldprd#1#2#3{{\rm Phys. ~Rev. ~}{\bf D#1}, #3 (19#2)}
\def\newprd#1#2#3{{\rm Phys. ~Rev. ~}{\bf D#1}, #3 (20#2)}
\def\plb#1#2#3{{\rm Phys. ~Lett. ~}{\bf B#1}, #3 (#2)}
\def\newplb#1#2#3{{\rm Phys. ~Lett. ~}{\bf B#1}, #3 (20#2)}
\def\npb#1#2#3{{\rm Nucl. ~Phys. ~}{\bf B#1}, #3 (19#2)}
\def\newnpb#1#2#3{{\rm Nucl. ~Phys. ~}{\bf B#1}, #3 (20#2)}
\def\prl#1#2#3{{\rm Phys. ~Rev. ~Lett. ~}{\bf #1}, #3 (19#2)}
\def\prl20#1#2#3{{\rm Phys. ~Rev. ~Lett. ~}{\bf #1}, #3 (20#2)}
\def\rep19#1#2#3{{\rm Phys. ~Rep. ~}{\bf #1}, #3 (19#2)}
\def\rep20#1#2#3{{\rm Phys. ~Rep. ~}{\bf #1}, #3 (20#2)}
\def\epjc#1#2#3{{\rm Eur. ~Phys. J.~}{\bf C#1}, #3 (#2)}

\bibliographystyle{unsrt}

\end{document}